\documentclass[11pt,letter, preprint]{article}
\usepackage{jheppub}
\usepackage{graphicx}
\usepackage{epsfig}
\usepackage{amsmath}
\usepackage{amsfonts}
\usepackage{amssymb}
\usepackage{url}
\usepackage{hyperref}
\usepackage{cancel,color}
\newcommand{\bea}{\begin{eqnarray}}
\newcommand{\eea}{\end{eqnarray}}

\newcommand{\be}{\begin{equation}}
\newcommand{\ee}{\end{equation}}
\newcommand{\br}{\begin{eqnarray}}
\newcommand{\er}{\end{eqnarray}}

\def\ptm{E{\!\!\!/}_T}
\def\beq{\begin{equation}}
\def\eeq{\end{equation}}
\def\bnq{\begin{eqnarray}}
\def\enq{\end{eqnarray}}
\def\barr{\begin{array}}
\def\earr{\end{array}}

\def\lsim{\buildrel{\scriptscriptstyle <}\over{\scriptscriptstyle\sim}}

\def\lapp{\mathrel{\rlap{\raise.5ex\hbox{$<$}}
                    {\lower.5ex\hbox{$\sim$}}}}
\def\gapp{\mathrel{\rlap{\raise.5ex\hbox{$>$}}
                    {\lower.5ex\hbox{$\sim$}}}}

\title{Dynamical R-parity Breaking at the LHC}
\author[a,b]{Shao-Long Chen,}
\author[c]{Dilip~Kumar~Ghosh,}
\author[a]{Rabindra N. Mohapatra,}
\author[d]{Yue Zhang}

\affiliation[a]{Maryland Center for Fundamental Physics and Department of Physics, \\
University of Maryland, College Park, Maryland 20742, USA}
\affiliation[b]{Institute of Particle Physics, Huazhong Normal University, Wuhan 430079, China}
\affiliation[c]{Department of Theoretical Physics, Indian Association for the Cultivation of Science, \\
2A\&2B Raja S.C.\,Mullick Road, Kolkata 700\,032, India}
\affiliation[d]{Abdus Salam International Centre for Theoretical Physics,  Strada Costiera 11, I-34014 Trieste, Italy}
\date{\today}

\abstract{In a class of extensions of the minimal supersymmetric
standard model with (B-L)/left-right symmetry that explains the
neutrino masses, breaking R-parity symmetry is an essential and
dynamical requirement for successful gauge symmetry breaking.
Two consequences of these models are: (i) a new kind of R-parity breaking interaction
that protects proton stability but adds new contributions to
neutrinoless double beta decay and (ii) an upper bound on the extra
gauge and parity symmetry breaking scale which is within the large hadron
collider~(LHC) energy range. We point out that an important
prediction of such theories is a potentially large mixing between
the right-handed charged lepton ($e^c$) and the superpartner of
the right-handed gauge boson ({\footnotesize$\widetilde W_R^+$}),
which leads to a brand new class of R-parity violating interactions of type
$\widetilde{\mu^c}^\dagger\!\nu_\mu^c e^c$ and $\widetilde{d^c}^\dagger\!u^c e^c$.
We analyze the relevant constraints on the sparticle mass spectrum
and the LHC signatures for the case with smuon/stau NLSP and gravitino LSP.
We note the ``smoking gun'' signals  for such models to be lepton
flavor/number violating processes: $pp\to \mu^\pm\mu^\pm e^+e^-jj$
(or $\tau^\pm\tau^\pm e^+e^-jj$) and $pp\to\mu^\pm e^\pm b \bar{b}
jj$ (or $\tau^\pm e^\pm b \bar{b} jj$) without significant missing
energy. The predicted multi-lepton final states and the flavor
structure make the model be distinguishable even in the early
running of the LHC.}

\emailAdd{chensl@iopp.ccnu.edu.cn}
\emailAdd{dilipghoshjal@gmail.com}
\emailAdd{rmohapat@umd.edu}
\emailAdd{yuezhang@ictp.it}

\arxivnumber{1011.2214}

\begin{document}
\maketitle

\section{Introduction}

Supersymmetry (SUSY) is one of the popular and best motivated candidates for
physics beyond the standard model (SM). It stabilizes the gauge
hierarchy and provides a dark matter candidate in a
natural manner. An intuitive requirement to stabilize the dark matter
in MSSM is the existence of the R-parity symmetry under which all
standard model particles are even and their superpartners are odd.
The lightest supersymmetric partner field (LSP), e.g., either the
neutralino or the gravitino, which is odd under R-parity is
therefore suitable as the dark matter candidate. If R-parity is
a global symmetry of the MSSM, it is logical to think of it as a
remnant of some high scale physics. It will of course be
interesting if the high scale physics is motivated by further
reasons. A shortcoming of R-parity conserving MSSM is the zero
neutrino mass. Understanding the origin neutrino masses then
requires it to be part of a larger theory. An example of extension
to the MSSM is to gauge the $B-L$ global symmetry, where anomaly
freedom requires introducing a right-handed neutrino to each
generation. The breaking of $B-L$ symmetry gives Majorana masses
to neutrinos. If the breaking is accomplished by Higgs fields with
$B-L=\pm 2$, it not only helps to explain the small neutrino
masses via the seesaw mechanism, but also leaves the R-parity as
an unbroken symmetry at the level of the MSSM~\cite{RNM}, thereby
providing a stable dark matter candidate. Extending MSSM by a
$B-L$ symmetry therefore ``kills two birds with one stone''.

Two possible classes of models with $B-L$ gauge symmetry are: (i)
$SU(2)_L\times U(1)_{Y}\times U(1)_{B-L}$, and (ii) its
left-right~(LR) symmetric generalization based on $SU(2)_L\times
SU(2)_R\times U(1)_{B-L}$. Breaking $B-L$ by two units in the
second case is more appealing since it can explain the origin parity violation,
and leads to a number of interesting phenomenological
implications for LHC searches including the $W_R$ boson as well for low
energy weak processes. We discuss them in this paper.

If the gauge symmetry is to be broken by a pair of Higgs
superfields $\Delta^c(1,3,+2)\oplus\bar\Delta^c(1,3,-2)$ which are
required to implement the seesaw mechanism and gauge anomaly
cancellation, two interesting results follow~\cite{kuchi}. First,
even though a priori the model is expected to have a remnant
R-parity after symmetry breaking, in its minimal version, exactly
the opposite happens, i.e., R-parity must be necessarily broken
spontaneously in order for the full gauge symmetry to break down
to the MSSM gauge group. If R-parity is exact, gauge symmetry
cannot break~\cite{kuchi}.
If the model is extended to include singlets, there is a range of
parameters where one can still have unbroken R-parity~\cite{babu}.
In the minimal model, however, R-parity breaking is mandatory.
%
%
The right-handed (RH) sneutrino field, $\tilde{\nu}^c$, has to
pick up a vacuum expectation value (VEV) along with the neutral
member of the $B-L=2$ triplet, breaking the parity symmetry and
contributing to the mass of the gauge bosons and gauginos
associated with right-handed currents. Since $\tilde{\nu}^c$ is an
R-odd particle, its VEV breaks R-parity. We call this class of
models ``{dynamical} R-parity breaking" models, since R-parity
breaking is forced on the theory at the global minimum of the
Hamiltonian. Other examples of models where R-parity breaking by
$\nu^c$ vev are the minimal $U(1)_{B-L}$ extensions of the
MSSM~\cite{Mohapatra:1986aw, barger}. In this note we will focus
on the SUSYLR case.

A consequence of dynamical R-parity breaking in  minimal SUSYLR
model is the prediction of an upper bound on the mass scale of the
right-handed $W_R$ boson, i.e., $M_{W_R}\lesssim M_{\rm
SUSY}/{f^2}$~\cite{kuchi2}, which is in the range accessible at
the LHC. Here $M_{\rm SUSY}$ is a generic soft SUSY breaking mass
scale, $f$ is the Yukawa coupling responsible for right-handed
neutrino masses and has to be $\gtrsim 0.1$. Similar relations are
also found in
 SUSYLR models where $\widetilde{\nu^c}$ vevs break
left-right symmetry\cite{Hayashi:1984rd,FileviezPerez:2008sx}.

Due to spontaneous R-parity breaking,
neutrino masses arise not only from the usual type-I seesaw
mechanism, but also via mixing with the neutralinos. Another
consequence is that neutralino is no longer a stable particle and
cannot therefore play the role of dark matter. However, if
gravitino is the LSP, it can have an extremely long lifetime
($\geq 10^{26}$ sec) and play the role of dark matter~\cite{Ji:2008cq}.
Implications for such a dark matter particle have been studied
extensively in connection with cosmic ray anomalies~\cite{ibarra}.


Since the scales of both superpartners and the new gauge interactions are predicted to lie in the
few TeV range, this theory could in principle be testable at the
hadron colliders~\cite{raidal}. In this paper therefore, we study the genuine signals from
dynamical R-parity breaking and discuss
how it can be distinguished from
usual R-parity breaking models at LHC.

We point out that the most important consequence of dynamical
R-parity breaking in SUSYLR models is a {\em large} mixing between
the RH charged lepton and RH wino, i.e., the physical RH charged
lepton after symmetry breaking is generically denoted as
\begin{equation}
\hat
\ell^c = \theta_{\ell\ell} \ell^c + \theta_{\ell W} \widetilde
W_R^+ + \cdots\,,
\end{equation}
where $\theta_{\ell\ell}, \theta_{\ell
W} \sim \mathcal{O}(1)$ for $\langle\widetilde \nu^c\rangle \simeq M_{\rm SUSY}
$ and the $\cdots$ represents the contributions of other
Higgsino fields if the corresponding Higgses VEV's also violate
parity. The physical charged lepton field contains a
large RH wino component and in turn induces new R-parity violating
terms of K\"{a}hler type.
This is characteristic of dynamical R-parity
breaking models~\cite{kuchi, Hayashi:1984rd, FileviezPerez:2008sx} with the presence of gauged $SU(2)_R$
and it leads to new effects absent in usual R-parity violating
MSSM or other models of spontaneous R-parity breaking,
such as~\cite{aulakh, masiero}. In particular, it leads to effective
R-parity violating interactions of the form $\widetilde{\mu^c}^\dagger\!\nu_\mu^c e^c$, $\widetilde{\tau^c}^\dagger\!\nu_\tau^c e^c$ and
$\widetilde{d^c}^\dagger\!u^c e^c$ for all generations of quarks/squarks. We show that
these kinds of vertices add new contributions to neutrinoless
double  beta decay and imply constraints on the parameters of the
model.

These new interactions bring rich phenomenology at the LHC.
In the context of a realistic model based on left-right symmetry, we study the
single production of a slepton NLSP from the RH neutrino decays,
which is produced via an on-shell $W_R$ boson
resonance at LHC. The NLSP single production and decay yield multi-lepton
final states of type
$pp\to \mu^\pm\mu^\pm e^+e^-jj$ (or $\tau^\pm\tau^\pm e^+e^-jj$)
and $pp\to \mu^\pm e^\pm b\bar{b}jj$ (or $pp\to \tau^\pm e^\pm b\bar{b}jj$)
which break both lepton number and flavor and have no
missing energy. The parent states could therefore be reconstructed
up to the oringinal $W_R$ decay. The lepton
final states are predicted to have distinct flavor structures.
We further point out that the $\ell^c -\widetilde W_R^+$ mixing
also leads to the production of righthanded polarized top quarks from down-type squark
decay, which is distinguishable from the $\lambda' QLd^c$ trilinear couplings in
the usual R-parity violating MSSM~\cite{huitu}.

In section II, we study the general features in a class of models where R-parity is broken
together with extra gauge symmetries.
We derive new R-parity breaking terms from the
K\"{a}hler potential and point out how to distinguish this class
of model from others, e.g. the MSSM with usual R-parity breaking
terms.
In section III, we review the symmetry breaking in the context of
minimal supersymmetric left-right (SUSYLR) model,
emphasizing the necessity of spontaneous (dynamical) R-parity violation
for $SU(2)_R$ gauge symmetry breaking.
We discuss the flavor issues of R-parity breaking and its implications
to neutrino mass and in section IV, we study the signatures of the model at the
LHC. We mainly focus on the single production and
decay of slepton NLSP via a heavier RH neutrino. The predicted
multi-lepton final states and the flavor structure make the model
distinguishable even in the early running of the LHC.
Finally in section V, we point out a new contribution
 to the neutrinoless double beta decay in the model and conclude.

\section{ Spontaneous R-parity Violation with Extended Gauge Symmetry}

Unlike explicit R-parity violation, spontaneous R-parity violation
(SRPV) has the advantage that it introduces only one new parameter
into the R-parity conserving theory -- the VEV of an R-parity-odd
field. Furthermore, if R-parity violating scale is at the TeV
range, above this temperature, R-parity is exact and therefore it
is less constrained by cosmology.

SRPV can be realized in various
ways: in the first model where the idea was
discussed~\cite{aulakh}, the superpartner of SM neutrino was given a
non-zero VEV. Since lepton number is not a gauge symmetry of the MSSM,
this leads to a doublet majoron which contributes to the $Z$-boson width and
LEP measurements therefore have ruled out this scenario. One could of
course implement SRPV by the VEV of a right-handed
sneutrino~\cite{masiero} in extensions of the MSSM that
explain neutrino masses. Since the right-handed sneutrino field is
a standard model singlet, the majoron does not couple to the $Z$-boson and therefore escapes the constraints
set by the LEP data.

In this section, we will pursue the implications when the R-parity is spontaneously broken
together with some extra gauge symmetry beyond $\bold{G}_{SM}=SU(2)_L\times U(1)_Y$.
%
%
%
Here we focus on the class of models where the extended gauge
group $\bold{G}$ contains a subgroup $SU(2)_R$. Clearly, the RH
neutrino and its superpartner will be charged under the $SU(2)_R$.
Giving a non-zero VEV to $\widetilde{\nu^c}$ will therefore give
rise to new interactions. Models with dynamical R-parity breaking
belong to this category. Furthermore, such model predicts that  the
scale of new gauge interactions is tied to the soft SUSY breaking
scale. This leads to several interesting new features as we show
below.

The key distinguishing prediction of such a model is the existence of a large
mixing between RH charged leptons and the gaugino superpartner of
the $W_R$ boson. Since we  work with the gauge group
$\bold{G}=SU(2)_L\times SU(2)_R\times U(1)_{B-L}$, the RH neutrino
and charged lepton form a doublet under $SU(2)_R$.

After the RH sneutrino developing a VEV
\begin{eqnarray}
\langle \widetilde L^c \rangle = \left[ \begin{array}{c}
\langle\widetilde{\nu}^c\rangle \\
0
\end{array} \right],
\end{eqnarray}
it breaks both the $SU(2)_R$ gauge group as well as the R-parity,
at the scale of $\langle\widetilde{\nu}^c\rangle$. Due to the
Higgs mechanism, the heavy gauge bosons acquire their mass by
absorbing the scalars ${\mathcal Im}\,\widetilde \nu^c$ and $\widetilde
\ell^c$ as the longitudinal components. Due to supersymmetry, one
would expect that a corresponding large Dirac mass would develop
between the $\nu^c-\widetilde Z'^0$ and $\ell^c-\widetilde W_R^+$.
Here we are interested in the chargino--lepton mixing, which is
the new source of R-parity breaking effects. We explicitly write
down the charged fermion mass matrix in the basis of $(\widetilde
W_R^+, \ell^{c+})-(\widetilde W_R^-, \ell^-)$,
\begin{eqnarray}\label{charge}
M_C=
\left[\begin{array}{cc}
M_{1/2} \ \ &  0 \\
M_{W_R} \ \ & m_\ell
\end{array}\right] \, ,
\end{eqnarray}
where $M_{W_R}=g_R\langle\widetilde\nu^c \rangle$ is the $W_R$
gauge boson mass and $M_{1/2}$ is the soft SUSY breaking mass for
the chargino. The (1-2) element is absent because neither the RH
neutrino nor the Higgs VEV couples $\ell^-$ and $\widetilde
W_R^+$. The determinant of this mass matrix is proportional to the
(light) charged lepton mass $m_\ell$ because the RH sneutrino VEV is an
electroweak singlet and therefore does not break chirality. This
means that there must be a physical state with the mass $m_\ell$
and identifiable as the charged lepton.

Diagonalization of this mass matrix leads to a mixing between
$\ell^c$ and $\widetilde W_R^+$. This mixing $\theta_{\ell W}$ is
large if  $M_{1/2}\simeq M_{W_R}$. The physical
charged lepton state is then given by
\be
\hat\ell^c =
\theta_{\ell\ell} \ell^c + \theta_{\ell W} \widetilde W_R^+\ ,
\ee
where $\theta_{\ell W}\sim \mathcal{O}(1)$. We note there is no such mixing
induced for $\widetilde W_R^-$.
Due to this mixing,
one can derive two new classes of R-parity violating interactions
from the right-handed gaugino matter coupling terms:
 First from the gaugino coupling $\sqrt2 g \widetilde W_R^+ {\nu_{\ell'}^c} \widetilde \ell'^{c\dagger}
+ {\rm h.c.} $, we get
\begin{eqnarray}\label{exotic1}
\mathcal{L}^{\ell'}_{\rm new} &=& \sqrt{2} g \theta_{\ell W} \left[ \hat\ell^c \nu_{\ell'}^c \widetilde {\ell'}^{c\dagger}  +
\bar {\hat\ell}^c \bar \nu_{\ell'}^c \widetilde {\ell'}^{c} \right] \ .
\end{eqnarray}
Analogously, using the right-handed gaugino interaction with
quarks and squarks,
one can also write down the new RPV
couplings for the squark-quark sector
\begin{eqnarray}\label{exotic2}
\mathcal{L}^q_{\rm new} &=& \sqrt{2} g \theta_{\ell W} \left[ \hat \ell^c u^c \widetilde d^{c\dagger}  +
\bar {\hat \ell}^c \bar u^c \widetilde d^{c} \right] \ .
\end{eqnarray}
Notice there is no supersymmetric counterpart of above terms generated,
unlike those from the superpotential: $\lambda LLe^c$, $\lambda' QLd^c$, etc..
The new couplings we obtain here break not only R-parity but also supersymmetry, since we started
from a mass matrix Eq.~(\ref{charge}) including the SUSY breaking gaugino mass.

As we will point out in Sec.~\ref{doublebeta}, the most stringent constraints on the
couplings in Eq.~(\ref{exotic2}) are from neutrinoless double beta
decay and HERA experiment~\cite{alla}, which tend to push the
squark and gluino masses to TeV. On the other hand, the LEP2
Z-pole observables give a universal constraint~\cite{Zpole} on
the mixing parameter $\theta_{\ell W}$ in both
Eqs.~(\ref{exotic1}) and (\ref{exotic2}), but it turns out to be
rather mild. Therefore, the sleptons masses are still allowed to
be not far above 100\,GeV.

\begin{figure}[hbt]
\begin{center}
\includegraphics[width=6cm]{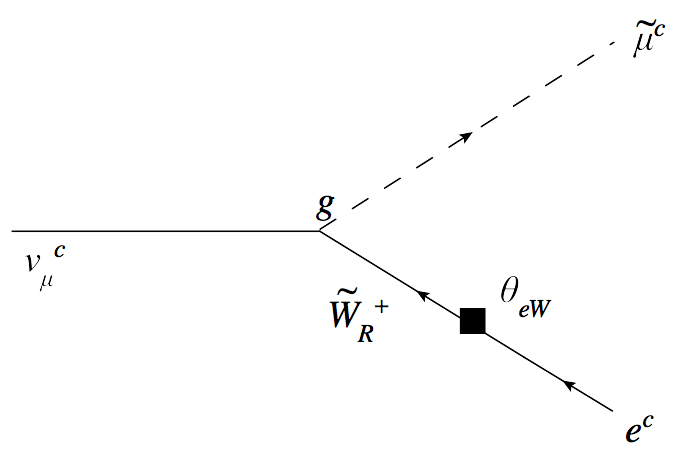}
\caption{Single production of RH slepton from RH sneutrino decays.
The black box represents RPV $e^c-\widetilde W_R$ mixing.}\label{newRPV}
\end{center}
\end{figure}

From these new interactions derived from dynamical R-parity breaking, one would expect the following
distinctive signatures at the LHC.
\begin{itemize}
\item Single production of slepton NLSP via Eq.~(\ref{exotic1}) and subsequent decays (see Fig.~\ref{newRPV}),
which is the main topic being studied in this paper.
\item Top quark produced from down-type squark decay via Eq.~(\ref{exotic2}),
whose polarization is opposite to the components from the SM background, as well as the conventional $\lambda'$ term.
\end{itemize}

We want to point out that these predictions are common to the
models with R-parity broken together with extended gauge symmetries,
as well as to those breaking $SU(2)_R$ symmetry
without the Higgs triplets, as long as the RH neutrino mass lies in the proper range.

\section{An Explicit Model of Dynamical R-parity Breaking}

In this section, we present a model based on the group
$SU(2)_L\times SU(2)_R\times U(1)_{B-L}$ where in the absence of
R-parity breaking, the gauge symmetry does not break\cite{kuchi}.
Thus the gauge dynamics dictates R-parity breaking. Hence it
explicitly provides an example for dynamical R-parity breaking
\footnote{For recent papers where hidden dynamics breaks R-parity,
see~\cite{jonas1} and also some experimental implications of such
models see~\cite{jonas2}. However, in these works, the R-parity is
not broken together with extended gauge symmetries that couple to
SM fermions, and therefore, does not predict the phenomenology of
dynamical R-parity breaking being discussed in this paper. }.

The model considered in this section does not have the discrete
parity symmetry. In the appendix, we will comment if such model
can be built in a completely parity symmetric form. Here, we first
review the salient features of the $SU(2)_L\times SU(2)_R\times
U(1)_{B-L}$  model (without parity symmetry) for completeness and
show how dynamical R-parity breaking occurs.

\subsection{SUSYLR: No gauge symmetry breaking without R-parity breaking}

The minimal SUSYLR model has the gauge group $G_{LR}=
SU(3)_c\times SU(2)_L\times SU(2)_R\times U(1)_{B-L}$. The
particle content and their representations under the gauge group
for the completely parity symmetric case are listed in Table
~\ref{content}. The $SU(2)_R$ Higgs triplets $\Delta^c$,
$\bar{\Delta}^c$ have been introduced to give mass to the RH
neutrinos and facilitate the seesaw mechanism. For the sake of
simplicity, we assume that the left-handed triplets $\Delta$,
$\bar{\Delta}$ are decoupled at high scale and do not exist in the
TeV theory.
The superpotential of the model is
\begin{eqnarray}
 W &=& Y_u Q^T \tau_2 \Phi_1 \tau_2 Q^c +  Y_d Q^T \tau_2 \Phi_2 \tau_2
Q^c \nonumber \\
 &+& Y_\nu L^T \tau_2 \Phi_1 \tau_2 L^c  + Y_l L^T \tau_2 \Phi_2 \tau_2
L^c +  i f \left(  L^{cT} \tau_2 \Delta^c L^c \right)  \nonumber \\
 &+& \mu_{\Phi\,ab} {\rm Tr} \left( \Phi_a^T \tau_2 \Phi_b \tau_2 \right)
+
 \mu_\Delta {\rm Tr} \left( \Delta^c \bar\Delta^c
\right) \ ,
\end{eqnarray}
where $Y$'s are Yukawa couplings, $f$ is the Majorana coupling of
leptons and $\mu_\Delta$ is the $\mu$-term for triplets. Note that
in the model there is no gauge singlets introduced.

\begin{table}[hbt]
\begin{center}
\begin{tabular}{ccccc}
\hline
  &\,$SU(3)_c$ &\,$SU(2)_L$ &\,$SU(2)_R$ &\,$U(1)_{B-L}$ \\
\hline
$Q$ & $\Box$ & $\Box$& {\bf 1} & $1/3$ \\
$Q^c$ & $\overline \Box$  & {\bf 1} & $\Box$ & $-1/3$ \\
$L$ & {\bf 1} & $\Box$ & {\bf 1} &$-1$ \\
$L^c$ & {\bf 1} & {\bf 1} & $\Box$ &$1$ \\
\hline
$\Phi_{1,2}$ & {\bf 1}  & $\Box$ & $\Box$  &$0$ \\
$\Delta^c$ & {\bf 1}  & {\bf 1} & $\Box$\hspace{-0.055cm}$\Box$  &$-2$ \\
$\bar \Delta^c$ & {\bf 1}  & {\bf 1} & $\Box$\hspace{-0.055cm}$\Box$  &$2$\\
\hline
$\Delta$ & {\bf 1}& $\Box$\hspace{-0.055cm}$\Box$ & {\bf 1}  &$2$ \\
$\bar\Delta$ & {\bf 1}& $\Box$\hspace{-0.055cm}$\Box$ & {\bf 1}  &$-2$ \\
\hline 
\end{tabular}
\caption[]{Particle content in the minimal SUSYLR model.
In this section, we first concentrate on the case with the left-handed Higgs triplets $\Delta$, $\bar \Delta$
do not exist in the TeV theory for simplicity. 
We will comment on the fully parity symmetric theory in the appendix.
}\label{content}
\end{center}
\end{table}

The corresponding soft terms are
\begin{eqnarray}
V_{\rm soft} &=& m_{\widetilde Q}^2 \left( \widetilde Q^\dag \widetilde Q +
\widetilde Q^{c\dag} \widetilde Q^c  \right) + m_l^2 \left( \widetilde L^\dag \widetilde
L + \widetilde L^{c \dag} \widetilde L^c \right) + m_{\Delta}^2  {~\rm Tr}(\Delta^{c\dag} \Delta^c)
+ m_{\bar\Delta}^2
  {~\rm Tr}(\bar\Delta^{c\dag} \bar\Delta^c)  \nonumber \\
& + &\frac{1}{2} \left(  M_{2L} \lambda_L^a \lambda^a_L + M_{2R}
\lambda_R^a \lambda^a_R + M_{1} \lambda_{BL} \lambda_{BL} + M_3 \lambda_g
\lambda_g \right)   \nonumber\\
& + & \widetilde Q^T \tau_2 A^q_i \phi_i \tau_2 \widetilde Q^c +
\widetilde L^T \tau_2 A^\ell_i \phi_i \tau_2  \widetilde L^c +
i A_f \widetilde
L^{cT} \tau_2 \Delta^c \widetilde L^c  \nonumber \\
& + &  B_{\Phi\,ab} {\rm Tr}\left( \tau_2 \phi_a^T \tau_2 \phi_b
\right) + B_\Delta  {\rm Tr}\left(  \Delta^c \bar\Delta^c \right)
  + {\rm h.c.} \ .
\end{eqnarray}
The D-term potential as well as the scalar potential can be found
in Refs.~\cite{kuchi, Ji:2008cq}.

The desired symmetry breaking pattern is $SU(2)_R\times
U(1)_{B-L}\to U(1)_Y$ at the first step, giving definite meaning
to the hypercharge $Y\,=\,I_{3_R}\,+\,{(B-L)/}{2}$, followed by the electroweak symmetry breaking.
The key point to note is that the potential does not break any gauge symmetry in supersymmetric limit~\cite{Aulakh:1998nn}.
Even if the soft SUSY breaking terms are included,
the gauge symmetry still remains unbroken as long as the RH sneutrino has zero VEV.
Parity and $SU(2)_R\times U(1)_{B-L}\to U(1)_Y$ breaking become possible only if the RH sneutrino picks up a
non-zero VEV.
The RH sneutrino being superpartner field has odd R-parity and therefore its VEV breaks
R-parity -- hence the claim~\cite{kuchi} that there is no parity breaking without R-parity breaking in the minimal SUSYLR model.
Furthermore, it was shown in~\cite{kuchi2} that RH sneutrino VEV is tied to the soft mass scale $M_{\rm SUSY}$.
This implies an upper bound on the $W_R$ gauge boson mass of order of the SUSY breaking scale.

To see this explicitly, we write down the potential including all VEV's given below;
\begin{eqnarray}\label{lrvevs}
\langle \widetilde L_e^c \rangle = \left[ \begin{array}{c}
\langle\widetilde{\nu}_{e}^c\rangle\\
0
\end{array} \right], \ \ \ \langle\Delta^c \rangle = \left[ \begin{array}{cc}
0 \ \ & 0 \\
v_R \ \ & 0
\end{array} \right], \ \ \
 \langle\bar \Delta^c\rangle = \left[ \begin{array}{cc}
0 \ \ & \bar v_R \\
0 \ \ & 0
\end{array} \right]\,,
\end{eqnarray}
where we choose to break the R-parity along the RH electron sneutrino $\widetilde{\nu}_{e}^c$ direction,
for phenomenological consideration to be explained in Section~\ref{3.2}.  The scalar potential involving the Higgs triplets and RH sneutrinos
is
\begin{eqnarray}
V &=& M_1^2 v_R^2 + M_2^2 \bar v_R^2 - 2 B v_R \bar v_R + |f|^2 \langle\widetilde\nu_e^{c}\rangle^4
+ \left[ 4 |f|^2 v_R^2 + m_0^2 - 2 |A| v_R - 2 |f| \mu_\Delta \bar v_R \rule{0cm}{4mm} \right]
\langle\widetilde\nu_e^{c}\rangle^2 \nonumber \\
&&+\ \frac{1}{8} (g^2+g'^2) (\langle\widetilde\nu_e^{c}\rangle^2-2v_R^2+2\bar v_R^2)^2 \ ,
\end{eqnarray}
where $M_1^2 = \mu_\Delta^2 + m_{\Delta}^2$, $M_2^2 = \mu_\Delta^2 + m_{\bar\Delta}^2$ and $B=B_\Delta$.
For simplicity, we have assumed the matrices $f$, $A_f$, $m_{\widetilde\ell}^2$
to be flavor diagonal and $f=f_{ee}$, $A=(A_{f})_{ee}$ and $m_0^2 = (m_{\widetilde\ell}^2)_{ee}$.
The VEVs of the Higgs bidoublets have been neglected in the first
stage of symmetry breaking. The potential should satisfy
$B<M_1M_2$ to be bounded from below. This can be seen by considering
D-flat directions $\langle\widetilde\nu_e^{c}\rangle=0$, $\langle \Delta\rangle = \langle \Delta^c \rangle = v \tau_1$
 and $\langle\bar \Delta\rangle = \langle \bar\Delta^c \rangle =\bar v \tau_1$,
 where $\tau_{1}$ is  the Pauli matrix.
 This scalar potential has the
property that on $\langle\widetilde\nu_e^{c}\rangle=0$ surface, there is no
symmetry breaking, i.e., $v_R=\bar v_R=0$ at the minimum. The
acceptable minimum that breaks parity and the gauge symmetries
therefore necessarily breaks R-parity.

\bigskip
\begin{figure}[t]
\begin{center}
\includegraphics[width=4.9cm]{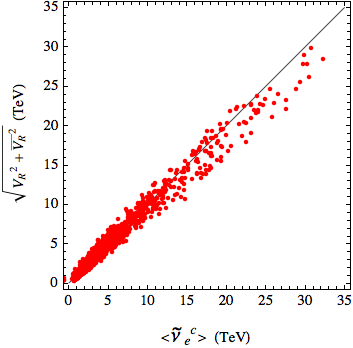} \;\;
\includegraphics[width=4.9cm]{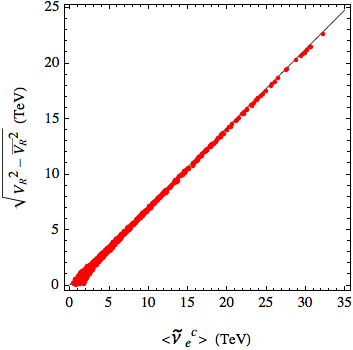}\;\;
\includegraphics[width=4.8cm]{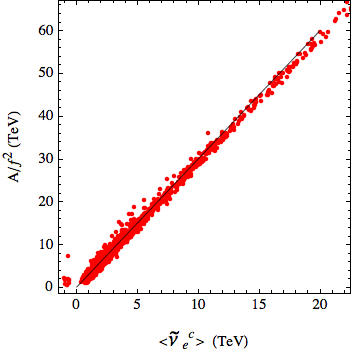}
\caption{Correlations among the VEVs.
The left-panel tells us that R-parity is broken as much as parity,
$\langle\widetilde\nu_e^c\rangle^2 \approx v_R^2+\bar v_R^2$. The middle panel shows
that the minimum always points towards the flat D-term potential
direction, $\langle\widetilde\nu_e^c\rangle^2 = 2(v_R^2-\bar v_R^2)$.
The right panel tells the value of the VEV is related to parameters in
the potential as $\langle\widetilde\nu_e^c\rangle\approx A/3f^2$. This agrees with the upper bound
obtained in Ref.~\cite{Ji:2008cq}.}\label{correlation}
\end{center}
\end{figure}

We have carried out numerical study of the minimization of the
scaler potential, by scanning over the bulk of parameter space
$(M_1, M_2, \sqrt{B}, A, \mu_\Delta, m_0) \in [100, 1000]\,$GeV
and $f \in [0.1, 0.5]$. The true vacuum must satisfy $V_{\rm
min}<0$ and $\langle\widetilde \nu_e^c\rangle\neq 0$. It turns out that there
are interesting correlations among the VEVs of RH neutrino and the
Higgs fields. They are shown in Fig.~\ref{correlation}. Typically,
we find the D-term potential always vanishes, i.e., $\langle\widetilde\nu_e^{c}\rangle^{2}
= 2(v_R^2-\bar v_R^2)$. Therefore, the physics at the RH scale
does not bring additional terms to the Higgs potential. The
sneutrino VEV and the Higgs triplets VEV's are of the same order,
$\langle\widetilde\nu_e^{c}\rangle^{2} \approx v_R^2+\bar v_R^2$, as well as an
approximate relation $\langle\widetilde\nu_e^c\rangle\approx A/3f^2$.
This agrees with the upper bound obtained in Ref.~\cite{Ji:2008cq}.
The key point is that, in SUSYLR model, the right-handed scale is
dynamically generated through the SUSY breaking soft mass
scale~\cite{kuchi},
\begin{eqnarray}
v_R \lesssim \frac{M_{\rm SUSY}}{f^2} \ ,
\end{eqnarray}
where $M_{\rm SUSY}\sim \mathcal{O}(100)$\,GeV corresponds
to the generic soft SUSY breaking mass scale.

\begin{figure}[hbt]
\begin{center}
\includegraphics[width=7.7cm]{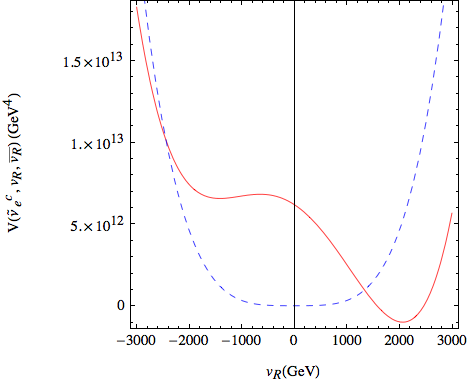} \hspace{0.5cm}
\includegraphics[width=7.1cm]{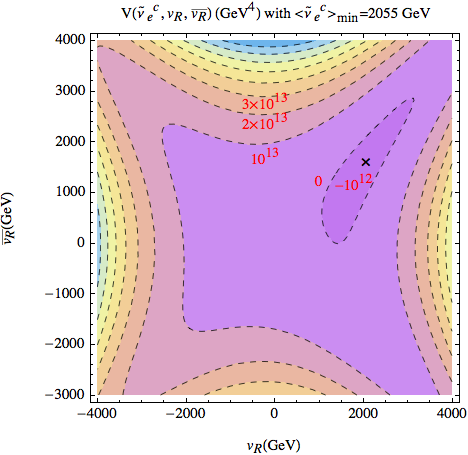}
\caption{Left panel: The potential $V$ as a function of $v_R$ for given
$\langle\widetilde\nu_e^c\rangle = 2055\,{\rm GeV}$,
$\bar v_R = 1607\,{\rm GeV}$ (solid curve) and
$\langle\widetilde\nu_e^c\rangle = \bar v_R = 0$ (dashed curve).
Right panel: Contour plot of the potential $V$ in the $v_R-\bar v_R$
plane for $\langle\widetilde\nu_e^c\rangle = 2055\,{\rm GeV}$.}\label{vevs}
\end{center}
\end{figure}

In order to illustrate the role of R-parity violation in symmetry
breaking, we choose the following set of parameters
\begin{eqnarray}
&M_1 = 213\,{\rm GeV}, M_2 = 251\,{\rm GeV}, \sqrt{ B } = 150\,{\rm GeV},& \nonumber \\
&\mu_\Delta = 517\,{\rm GeV},
 A = 240\,{\rm GeV}, m_0 = 376\,{\rm GeV}, f=0.21 & \, .
\end{eqnarray}
The resulting VEV's and the minimum potential value are
\begin{eqnarray}
&&\langle\widetilde\nu_e^c\rangle = 2055\,{\rm GeV},\ \ v_R = 2063\,{\rm GeV}, \ \  \bar v_R = 1607\,{\rm GeV},  \ \  V_{\rm min} = -1.0\times 10^{12}\,{\rm GeV}^4 \ .
\end{eqnarray}
The configuration of the potential around the vacuum is shown in
Fig.~\ref{vevs}. Clearly, the global minimum of the potential
breaks R-parity, i.e., $\langle\widetilde\nu^c_e\rangle\neq0$.
Because the R-parity and the lepton number are broken simultaneously
with the gauge symmetries, no massless Majoron is present. On the other hand,
the dynamical R-parity breaking associated with gauge symmetry
breaking at few TeV scale offers rich phenomenology.

\subsection{Neutrino mass and flavor alignment of R-parity violation}\label{3.2}

Neutrino masses in this model have been discussed extensively in~\cite{Ji:2008cq}.
We review the salient points for completeness and, in particular,
constraints on flavor of the R-parity violation.
In SUSYLR model, the matter fields obtain
their Dirac masses from the coupling to the Higgs bidoublets
$\Phi_{1, 2}$. Generally, there are four $SU(2)_L$ Higgs doublets
at the electroweak scale. The additional neutral Higgs bosons will
lead to flavor changing neutral currents at tree level. This can
be suppressed either by proper doublet-doublet splitting or by
cancellations~\cite{Zhang:2007qma}. Another way to avoid the flavor changing Higgs
effects is by replacing the second $B-L=0$ bidoublet with a $B-L=2$ bidoublet
as has recently been suggested~\cite{diego}. We do not discuss this
here.

We assume that the bidoublet Higgs fields take the following form of VEV's,
\begin{eqnarray}
\Phi_1 = \left[ \begin{array}{cc}
0 \ \ & 0 \\
0 \ \ & \kappa_1
\end{array} \right], \ \ \Phi_2 = \left[ \begin{array}{cc}
\kappa_2 & \ \ 0 \\
0 & \ \ 0
\end{array} \right] \ ,
\end{eqnarray}
where $\kappa_1$ gives Dirac masses to the up-type quarks and
neutrinos, while $\kappa_2$ contributes to down-type quarks and
charged leptons masses. With this VEV structure, the $W_L-W_R$ gauge bosons do not mix with each other.
 Here we mainly focus on the lepton sector.
We will attribute the hierarchies among the charged lepton masses
to the Yukawa couplings. In particular, we focus on the
low $\tan\beta=\kappa_2/\kappa_1 \sim \mathcal{O}(1)$ regime. In
this case, the Yukawa coupling constants are set as $y_\tau \approx
10^{-2}$, $y_{\mu} \approx 10^{-3}$, $y_e \approx 10^{-5.5}$ and
$(Y_\nu)_{ij}\approx 10^{-6}$.
Even though there are
four SM Higgs doublets (or two MSSM Higgs pairs), since only two
of them contribute to fermion masses and the other two play the role
of spectators, our tan$\beta$ is same as the MSSM one.
The $\mu_\Phi$ and $B_{\Phi}$ play a similar role as the $\mu$ and $B_\mu$
parameters in the MSSM for the electroweak symmetry breaking.

\begin{figure}[hbt]
\begin{center}
\includegraphics[width=6cm]{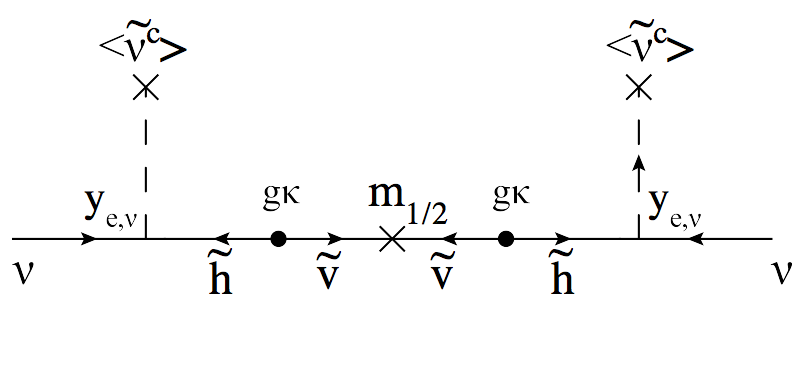}\hspace{1cm}
\includegraphics[width=5.5cm]{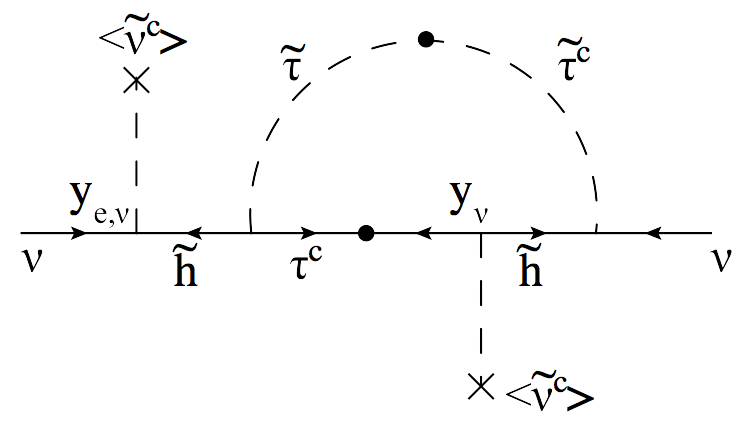}
\caption{Contributions to neutrino masses from R-parity violation. Left panel: tree-level contribution
due to neutrino-neutralino mixing. Right panel: loop-suppressed radiative correction to the neutrino masses.
The $\tilde{\rm v}$ represents neutral gaugino fields. The black dots are the usual Higgs VEV insertions.}\label{neutrino}
\end{center}
\end{figure}

Because of R-parity violation, there are additional contributions to neutrino masses,
on top of the Type-I seesaw mechanism. They arise from neutrino-neutralino
mixing which has been calculated in Ref.~\cite{Ji:2008cq} with R-parity
breaking in the RH electron sneutrino direction,
\begin{eqnarray}\label{numass}
 \left(M_{\nu}\right)_{ij} &\approx&- \frac{g_L^2 g'^{2} \kappa_2^2}{2 M_{2L} M_{\widetilde B}
( \mu_{11}\mu_{22} - \mu_{12}^2)} \left(\frac{M_{2L}}{g_L^2} +
\frac{M_{2R}}{g_R^2} + \frac{M_{1}}{g_{\rm BL}^2}\right)  \\
&\times& \left[ \left( \frac{\mu_{12}}{\mu_{11}} \right) (Y_\nu)_{i1}
(Y_\nu)_{1j} +  (Y_\nu)_{i1} y_e \delta_{j1} + (Y_\nu)_{1j}
y_e \delta_{i1} - \left( \frac{\mu_{11}}{\mu_{12}} \right) y_e^2
\delta_{i1} \delta_{j1} \right]\langle\widetilde\nu^{c}_e\rangle^2 \nonumber \ ,
\end{eqnarray}
assuming the LH sneutrino VEV's are negligible.
$M_{L,R}$ and $M_{BL}$ are soft supersymmetry breaking gaugino masses.
The corresponding Feynman diagram is shown in the left panel of Fig.~\ref{neutrino}.
Choosing the R-parity violation along the
$\widetilde\nu^c_e$ direction helps to avoid too large
contributions through the $y_\tau$, $y_\mu$ couplings, while the
electron and neutrino Yukawa couplings are sufficiently small to
keep the neutrino mass scale $\mathcal{O}(0.1)\,$eV in tact.

We also notice that there are radiative corrections to the
neutrino mass~\cite{Barbier:2004ez, Ba2}, which are also proportional to the corresponding Yukawa coupling and are loop suppressed
(see the right panel of Fig.~\ref{neutrino}, as well as Fig.~2 in~\cite{goran}).
So they are safely small as long as $\langle\widetilde\nu^c_\mu\rangle$ and $\langle\widetilde\nu^c_\tau\rangle$ are
vanishing and $\tan\beta$ is low.

\bigskip
{\it The above discussions justify our choices of VEV configuration in Eq.~(\ref{lrvevs}). }

\bigskip
Before closing this section, we comment on the flavor violations.
The next section will mainly concentrate on the scenario with slepton NLSP singly
produced from the $W_R^\pm$ gauge boson resonance,
hence we need to understand the existing experimental constraints on the relevant mass scales.
In the minimal non-SUSY left-right model, the famous
neutral $K$-meson mixing tends to push to $W_R$ mass to be above 2.4\,TeV~\cite{Zhang:2007fn}.
In the supersymmetric version, loop diagrams mediated by
superpartners also make additional contribution to both quark and lepton flavor violation processes.
They are safely small if the relevant mass scales are high enough.
Otherwise, in order to optimize the discovery prospects at the LHC,
the superpartners and $W_R$-boson masses have to lie in the (sub-)TeV regime,
which requires fine-tuning the flavor structures of the model in a similar way to the MSSM.
This is nothing but the SUSY flavor problem, and the constraints on scales are quite model dependent.
In principle, there could also be contribution to flavor violations from higher dimensional operators
controlled by unknown physics in the UV.

Therefore, in the following we shall only adopt the bounds
$M_{W_R}> 1\,$TeV and $m_{\widetilde \ell}> 100\,$GeV
from Tevatron and LEP2 direct searches, respectively.

\section{Single Production of Slepton NLSP and its Decay at the LHC}
In this section, we start exploring the LHC implications of this
model. First we need to know the R-parity violating (RPV)
interactions that induce the decays of the sparticles produced.

\subsection{Relevant RPV couplings}

In general, spontaneous R-parity breaking through the RH sneutrino VEV
generates the bilinear terms in the superpotential and the soft
potnetial
\begin{eqnarray}
W_{\cancel{R}} = \mu_i L_i H_u \ ,  \ \ \ \ \
V_{\cancel{R}-\rm soft} = B_i \widetilde L_i H_u + {\rm h.c.} \ .
\end{eqnarray}
The bilinear
term facilitates the R-parity breaking decay of the lightest
neutralino, $\widetilde\chi_1^0$ as follows:
$\widetilde\chi_1^0\to Z^0\nu$, $\widetilde\chi_1^0\to
W^\pm\ell_\mp$ or $\widetilde\chi_1^0\to
\ell_1^+\ell_2^-\nu$~\cite{barger}.
In the literature, more complete collider phenomenologies of R-parity violation from the
superpotential has been studied in detail and reviewed in
Refs.~\cite{Barger:1989rk, Allanach:1999bf, deCampos:2007bn, Barbier:2004ez}.

In the SUSYLR model, the bilinears arise from the electron and neutrino Yukawa couplings
and the corresponding $A$-term once the RH sneutrino VEV is inserted~\cite{Ji:2008cq}.
Therefore
\begin{eqnarray}
\frac{\mu_i}{\mu_\Phi} \simeq \frac{B_i}{B_\Phi} \simeq y_e,\,y_{\nu} \simeq 10^{-6} \ .
\end{eqnarray}
These bilinear terms will induce trilinear R-parity breaking terms $\lambda LLe^c$, $\lambda' Q L d^c$.
The most important terms for the following study are those associated with third generation fermions,
\begin{eqnarray}\label{trilinear}
\lambda'_i t \ell_i b^c + \lambda_i \nu \ell_i \tau^c \ ,
\end{eqnarray}
where $\lambda'_i = y_t \mu_i/\mu_\Phi$, $\lambda_i = y_\tau \mu_i/\mu_\Phi$ and $i=1,2,3$.
On the other hand, the $\lambda''$ term will not be generated, since baryon number symmetry is
respected by the sneutrino VEV, thereby guarantee the proton stability.

\bigskip
As already stated in previous sections, a distinct feature that arises when R-parity
is dynamically broken together with an $SU(2)_R$ gauge symmetry
is the large mixing between the RH electron $e^c$ and
the $SU(2)_R$ gaugino, i.e., $\widetilde{W}^+_R$.
This is not present in the MSSM with general R-parity violation.
The form of charged fermion mass matrix and
the obtained mixing in SUSYLR model is given explicitly in the Appendix.
From the usual gaugino Yukawa-like coupling term for $\mu$
and $\tau$ flavors,
one obtains the new couplings (similar to Eq.~(\ref{exotic1}))
\begin{eqnarray}\label{exotic}
\mathcal{L}^\ell_{\rm new} &=& \sqrt{2} g \theta_{eW} \left[ e^c \nu_\mu^c \widetilde \mu^{c\dagger}  +
\bar e^c \bar \nu_\mu^c \widetilde \mu^{c} \right] + \sqrt{2} g \theta_{eW} \left[ e^c \nu_\tau^c \widetilde \tau^{c\dagger}  +
\bar e^c \bar \nu_\tau^c \widetilde \tau^{c} \right] \,.
\end{eqnarray}
As we will see, these interactions open new channel for the single production of a slepton at hadron colliders (Fig.~\ref{newRPV}).
Similarly, from the neutralino mass matrix, one can also obtain large mixing between the $\widetilde Z'$ and $\nu_e^c$,
and in turn the  couplings
\begin{eqnarray}
\mathcal{L}^{\nu^c}_{\rm new} = \sqrt{2}g_{_{Z'}} \theta_{_{NZ'}} \left( \nu^c_e \mu^c \widetilde \mu^{c\dagger} +
 \nu^c_e \tau^c \widetilde \tau^{c\dagger} + \nu^c_e \nu^c_\mu \widetilde \nu^{c\dagger}_\mu
+ \nu^c_e \nu^c_\tau \widetilde \nu^{c\dagger}_\tau \right) + \rm h.c..
\end{eqnarray}
In principle, sparticle single production could also happen through the mixing between
$\nu_e^c$ and $\widetilde Z'$~\cite{goran} which, however, calls for some tuning between $M_{Z'}$ and $M_{1/2}$.

Contrary to the usual R-parity breaking term from superpotential,
these new R-parity breaking sources come from the
gaugino Yukawa-like couplings (in the K\"{a}hler potential). As we illustrate in the below,
such theories could be tested at the LHC where the new gauge interactions are accessible.

\subsection{Branching ratios of slepton NLSP decay}

From the previous sections one learns that, in the SUSYLR model
under discussion,  a new class of RPV couplings
Eq.~(\ref{exotic}) emerge due to the mixing between $e^c$ and
$\widetilde W_R^+$. To study its implications for hadron colliders, we
need to know the sparticle spectrum.
If one takes the assumption of universal scalar masses at high scale,
the RH sleptons are likely to be the lightest among matter superpartners
in the MSSM due to the smaller Yukawa couplings as well as
the smaller weak gauge couplings~\cite{book}. The situation would be
similar in SUSYLR
models. The lightest slepton could be a stau or the smuon
depending on detailed parameter range. In our study, we will
assume that smuon is the lightest superpartner above the gravitino, the latter
in our model could be the very weakly unstable dark matter.

As promised, we study the implications of scenario at the LHC for the case
where smuon or stau is the NLSP among the superpartners. Due to
the relatively low tagging efficiency of the tau lepton, we would
focus on the smuon.

\begin{figure}[hbt]
\begin{center}
\includegraphics[width=7.2cm]{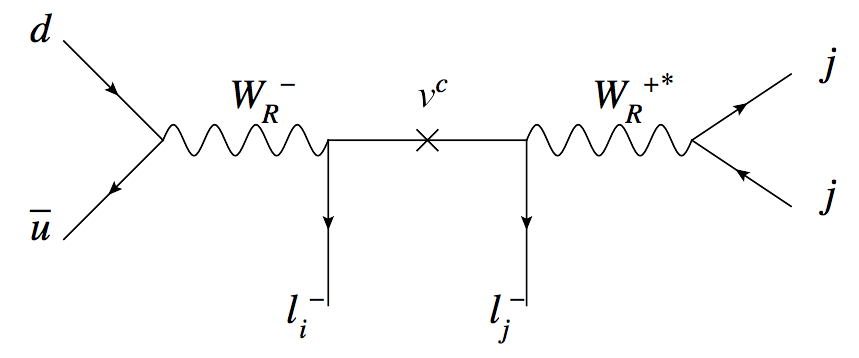}\hspace{1cm}
\includegraphics[width=6.4cm]{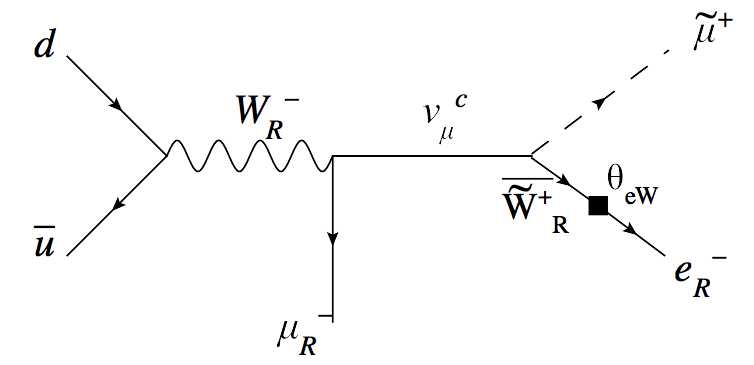}
\caption{Production of the RH neutrino and its decays.
Left-panel: the usual same-sign lepton diagram for $W_R$ discovery. Right-panel:
Single production of the smuon through RH neutrino RPV decay. In this case,
there is equal possibility to break the muon lepton number twice through the Majorana mass of
$\nu_\mu^c$ or not, so one can get either $\widetilde \mu^+ e_R^-$ or $\widetilde\mu^- e_R^+$ from its decay.
The black box
represents sneutrino VEV insertion as indicated in Fig.~\ref{newRPV}. }\label{compete1}
\end{center}
\end{figure}

The new LHC signals originate from the production of $W_R$ in $pp$
collision and its subsequent decay to muon and RH muon neutrino
which subsequently decays. In the non-SUSY LR models with type I seesaw the RH neutrino decays
mostly to the three body final state $\ell^\pm \ell^\pm jj$ via $W_R$ exchange~\cite{Keung:1983uu, Gninenko:2006br}. However in the SUSY version, if
the smuon, $\widetilde\mu$ is lighter than $\nu^c$, an interesting new two body final state channel emerges:
RH neutrino decays to a $\widetilde\mu$ and an electron. Since this is two-body decay, for the smuon
sufficiently light, it will certainly dominate over the three body non-SUSY mode,
\footnote[2]{Since the RH neutrino mass matrix is
proportional to the matrix $f$, which we have taken to be
diagonal in the basis of physical charged leptons,
there is no further flavor changing in $\nu^c$ mass matrix (propagator).}
as shown in
Fig.~\ref{compete1}. Therefore, the smuon single production could
take this advantage  and be large enough to be probed at the LHC.

\begin{figure}[hbt]
\begin{center}
\includegraphics[width=5cm]{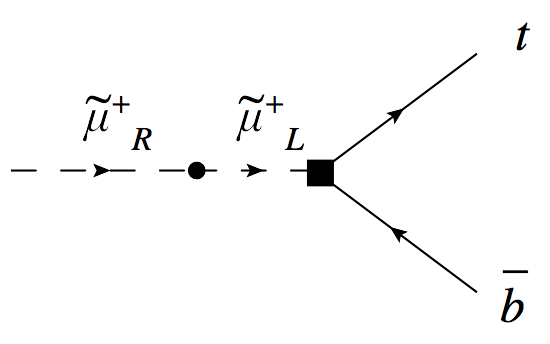}\hspace{1cm}
\includegraphics[width=7cm]{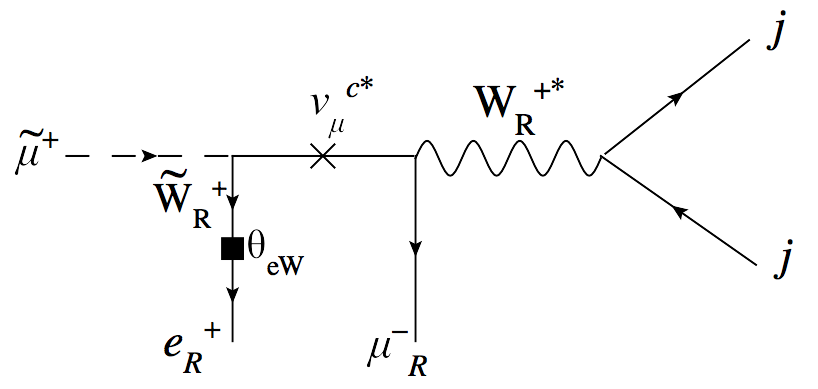}
\caption{Two- and four-body decay modes of the smuon NLSP. The black box
represents sneutrino VEV insertion as indicated in Fig.~\ref{newRPV}.
The black dot stands for the usual Higgs VEV insertions.
Hereafter, we denote the RH smuon as $\widetilde \mu^c \equiv \widetilde \mu_R^+$.}\label{decay1}
\end{center}
\end{figure}

As the NLSP, the smuon $\widetilde \mu_{\rm _{NLSP}} = \cos \alpha\ \widetilde \mu_R + \sin \alpha\,e^{i\beta}
\widetilde \mu_L$ will decay dominantly through R-parity
breaking interactions rather than the Planck scale suppressed
decay to the gravitino.
In the case where there is a large mixing between LH and RH smuons, i.e., $\sin\alpha \sim \mathcal{O}(1)$,
$\widetilde \mu^+$ can decay to $t\bar b$ or $\tau \bar\nu$ through the induced trilinear RPV terms as shown in Eq.~(\ref{trilinear}),
although suppressed by the small $y_e$ or $y_\nu$ Yukawa couplings.
Note that electroweak symmetry forbids the direct coupling of RH
smuon to $\bar f f$, and the RH sneutrino VEV does not help
because it is also a singlet. Only the LH and RH smuon mixing term
$(m_{\widetilde \mu}^2)_{LR}$ which is proportional to the Higgs
doublet VEV, can facilitate this decay.

\begin{figure}[b]
\begin{center}
\includegraphics[width=6.9cm]{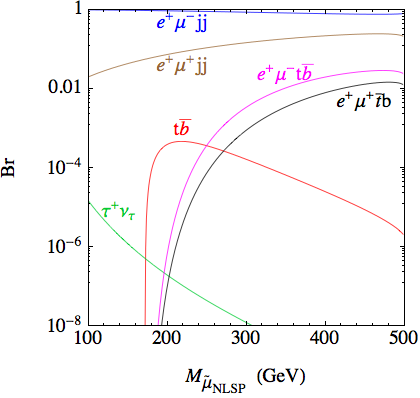}\ \ \
\includegraphics[width=7.0cm]{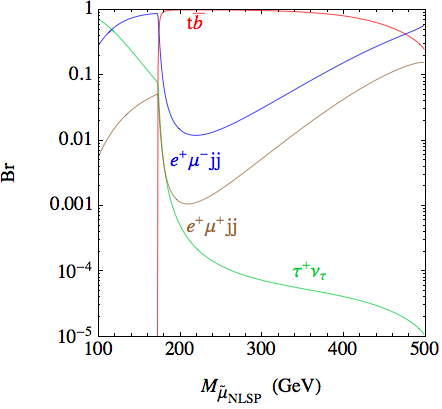}
\caption{Branching ratios for the smuon NLSP decay. The left panel represents the
suppressed LH and RH slepton mixings ($\sim10^{-3}$) case, while in the right panel,
we take an unsuppressed $\mathcal{O}(1)$ such mixing.
Charge conjugated final states are not listed but also possible.}\label{branching1}
\end{center}
\end{figure}

On the other hand, if the mixing term $(m_{\widetilde
\mu}^2)_{LR}$ is severely suppressed, i.e., $\sin\alpha \ll 1$, the smuon is almost purely RH.
Therefore, it decays through a four-body channel, as shown in the right
panel of Fig.~\ref{decay1}.
Such decay rate is proportional to the
gauge coupling instead of the small $y_e$ or $y_\nu$ couplings.
It could be comparable or even dominate over the above two-body
decays when the latter is further suppressed by the LR smuon mixing.
In principle, $\widetilde\mu^c$ can decay to both $e^+\mu^\pm jj$ final states. However, since the
intermediate RH neutrino is off-shell, the probablity to break the muon lepton number is larger than
that conserving the lepton number. This point can be see from the blue and brown curves in the left panel of Fig.~\ref{branching1}.

The branching ratios to different final states of the smuon decay
has been plotted in Fig.~\ref{branching1} with the following parameters
chosen, $M_{W_R}=2\,$TeV, $M_{\nu^c_\mu}= 500\,$GeV and $g \theta_{eW} = 0.2$.
In the suppressed LR slepton mixing case (left panel, here and below, we will take $\sin\alpha \approx 10^{-3}$
as a benchmark point), the four-body decay of smuon NLSP always dominates over all the two-body channels.
In the large mixing case (right panel), the $t\bar b$ channel will dominate if it is
kinematically allowed, while $\tau \bar\nu$ and the four-body channels $e^+\mu^-jj$ could
respectively dominate in certain low smuon mass windows.
In both cases, since the Majorana RH neutrinos are involved in the production
and/or decay processes, lepton number can be broken, which leads to the most
promising discovery channels at the LHC.
The expected signatures are listed in Table~\ref{table2}.

\begin{table}[thb]
\begin{center}
\begin{tabular}{|c|c|c|}
\hline
&~large $\widetilde\mu-\widetilde\mu^c$ mixing~&~suppressed mixing  \\
\hline
$M_{\widetilde\mu^c}>m_{t}+m_b$~& $pp\to \mu^- e^- t\bar b,\ \mu^+ e^+ \bar t b$~&  \\
\cline{1-2}
  &$pp\to \mu^\pm \mu^\pm e^+e^-jj$~& \\
\raisebox{3ex}[0pt]{$M_{\widetilde\mu^c}<m_{t}+m_b$} & $pp\to \mu^-e^+\tau^++\cancel{E}_T$~&~
\raisebox{4.7ex}[0pt]{$pp\to \mu^\pm \mu^\pm e^+e^-jj$} \\
\hline
\end{tabular}
\end{center}
\caption[]{Expected final states in the single production of the NLSP $\widetilde\mu^c$
with $M_{\widetilde\mu^c}<M_{\nu^c}$ assumed. Large and suppressed $\widetilde\mu-\widetilde\mu^c$
mixing cases are both listed.}\label{table2}
\end{table}

In the following two subsections we will discuss the signature of the single
production and decay of smuon NLSP via a heavier right-handed neutrino. We
also discuss possible standard model backgrounds and elaborate on the
selection criteria necessary for such signals to be significantly observed
over the standard model background. The large number of diagrams involved
in the standard model background processes are calculated using the helicity
amplitude package MadGraph~\cite{madgraph} and CalcHEP 2.5.4~\cite{pukhov}.
To estimate the number of signal and background events as well as their phase
space distribution(s), we use a parton-level Monte-Carlo event generator.
In our numerical analysis, we use the CTEQ6L parton distribution function
\cite{cteq6l} and fix the factorization scale $Q^2 = \hat{s}/4 $.
In our parton-level simulation of both signal and background events,
we smear the leptons and jet energies with a Gaussian distribution according to
\beq
\frac{\delta E }{E} = \frac{a}{\sqrt{E/{\rm GeV}}} \oplus~b
\label{gaussian_smear}
\eeq
with the CMS parameterization, $a_{\ell} = 5\%, b_{\ell} = 0.55\%$ and
$a_j = 100\%, b_j = 5\%$, $\oplus $ denotes a sum in quadrature.

\subsection{$pp\to e^+e^-\mu^\pm \mu^\pm jj$}

This particular final state dominates when the mixing between LH and RH smuon
is suppressed (or in the low mass ($\lsim M_{\rm top}$) region for a
large mixing). In this section, we will denote smuon NLSP as $\widetilde \mu^c$ since
it is mainly the RH component.
The most striking feature of this final state is the {\it
three same sign leptons } and one opposite sign lepton associated
with two jets without missing energy.
Assuming the narrow width approximation for $\nu^c $ and $\widetilde \mu^c$,
we can simply write down the signal cross-section
$ \sigma_s (pp\to e^+e^-\mu^+\mu^+ jj)$ as
\bnq
&&\sigma (pp\to e^+e^-\mu^+\mu^+ jj) \approx
\sigma ( pp \to W^{+}_R \to \mu^+ \nu^c_{\mu}) \\
&&\hspace{1.5cm} \times \left[ {\rm Br}(\nu^c_{\mu} \textcolor[rgb]{1,0,0}{\to} \tilde\mu^c e^-)
 \times {\rm Br}(\tilde\mu^c \to e^+ \mu^c jj)  +  {\rm Br}(\nu^c_{\mu} \to \tilde\mu^{c\dagger} e^+)
 \times {\rm Br}(\tilde\mu^{c\dagger} \textcolor[rgb]{1,0,0}{\to} e^- \mu^+ jj)  \rule{0cm}{4mm}\right] \ ,  \nonumber
\label{csA_approx1}
\enq
where the red (dotted) arrow indicates lepton number violation by two units
on the involved RH neutrino propagator.
The charge conjugated final state $ \sigma (pp\to e^+e^-\mu^-\mu^- jj) $ which
is mediated by the intermediate $W^{-}_R$ boson can be similarly approximated.
In our analysis, we combine both these two final states.
We define the signal identification with four charged leptons and two jets.
The events are further selected by the following
set of cuts
 \begin{enumerate}
 \item We require that both jets and leptons should appear
within the detector's rapidity coverage, namely
\beq
  |\eta(\ell)| <\ 2.5,  \ \ \  |\eta(j)| <\ 3 \ .
     \label{cut:eta}
\eeq

\item
The leptons are ordered  according to their transverse momentum $(p_T)$ hardness and
the $p_T$ of the leading lepton must satisfy
\beq
p_T(\ell_1) > 100~{\rm GeV} \ ,
\label{ptlhard_cut}
\eeq
and for rest of the leptons
\beq
p_T(\ell) > 15~{\rm GeV} \ .
\label{ptl_cut}
\eeq

For two associated jets we demand that
\beq
p_T^{{\rm jets}} > 25~{\rm GeV}\,.
\label{ptj_cut}
\eeq

\item We must also ensure that the jets and leptons are well
  separated so that they can be identified as individual entities. To
  this end, we use the well-known cone algorithm defined in terms of a
  cone angle $\Delta R_{\alpha\beta} \equiv \sqrt{ \left
  (\Delta \phi_{\alpha\beta} \right)^2 + \left (\Delta
  \eta_{\alpha\beta} \right )^2} $ with $\Delta \phi$ and $\Delta \eta$
  being the azimuthal angular separation and rapidity difference
  between two particles. We demand that
  \beq
    \Delta R_{jj} > 0.4 \ , \quad
    \Delta R_{\ell j} > 0.4 \ ,\quad
     \Delta R_{\ell \ell}> 0.4 \ .
     \label{deltar_cut}
   \eeq

\item In our analysis,
We use simplified definition for the missing transverse energy:
$\ptm = \sqrt {\left ( \sum p_x\right )^2 + \left ( \sum p_y \right )^2 }$,
where the sum goes over all observed charged leptons and jets.
We demand that there is no significant missing energy in our signal
\beq
\ptm < 30~{\rm GeV}.
     \label{ptm_cut}
\eeq
\end{enumerate}

\begin{figure}[t!]
\begin{center}
\includegraphics[width=7.9cm]{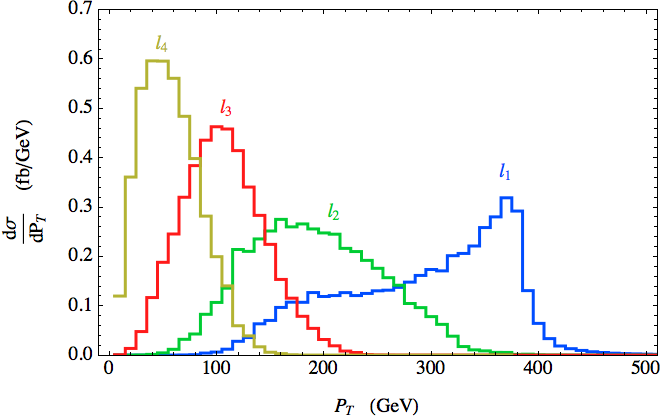}
\caption{$p_T$ distributions of all four leptons in the process
$pp\to e^+e^-\mu^\pm\mu^\pm jj$ at $\sqrt{s} = 14 $ TeV.
The leptons are ordered according to their $p_T$ hardness
$(p_T(\ell_1) > p_T(\ell_2) > p_T(\ell_3) > p_T(\ell_4) )$.  We have
fixed $M_{W_R} =1 $ TeV, $M_{\nu^c_\mu} = 500 $ GeV, and
$M_{\tilde \mu^c} = 300 $ GeV. }
\label{distn}
\end{center}
\end{figure}

\begin{figure}[thb]
\begin{center}
\includegraphics[width=4.9cm]{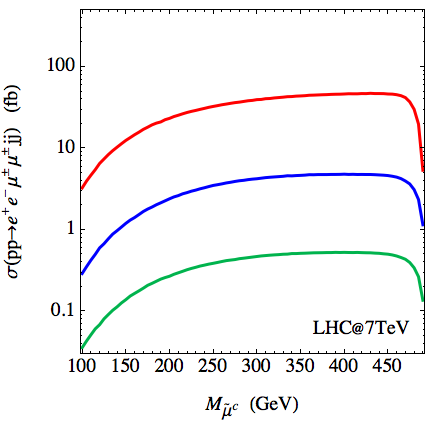}\ \ \
\includegraphics[width=4.9cm]{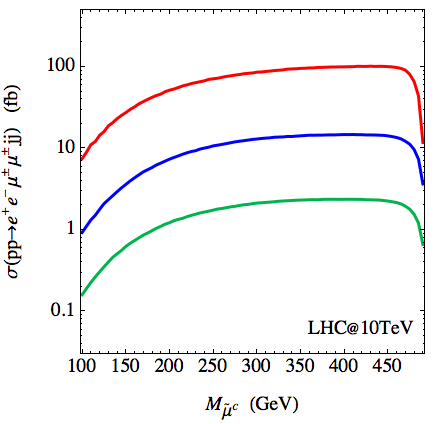}\ \ \
\includegraphics[width=4.9cm]{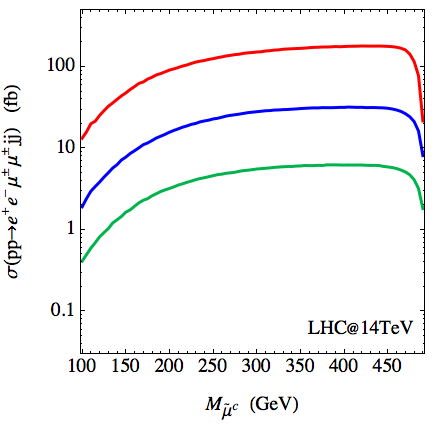}
\caption{Signal cross sections $\sigma (pp\to e^+e^-\mu^\pm\mu^\pm jj)$
(after all cuts as mentioned in the text) as a function of smuon mass at
the LHC with $\sqrt{s} = 7$ TeV, 10 TeV and 14 TeV. Three curves from
top to bottom in each panel correspond to $M_{W_R} =1 $ TeV,
1.5 TeV and 2 TeV respectively. $M_{\nu^c_R} $ is kept fixed at 500 GeV. }
\label{Xs_A}
\end{center}
\end{figure}

Our choice of $p_T$ cut on the leading lepton  (Eq.~(\ref{ptlhard_cut}))
can be well justified from the $p_T$ distribution of all four
leptons as displayed in Fig.~\ref{distn} assuming
$M_{W_R} =1$\,TeV, $M_{\nu_\mu^c} = 500$\,GeV and $M_{\tilde\mu^c} = 300$\,GeV
and at $\sqrt{s} = 14$\,TeV. Here, one should note that while generating
$p_T$ distributions (Fig.~\ref{distn}), we impose an uniform
loose cut $(p_T > 15~{\rm GeV})$  on all four leptons, however,
rest of the cuts remain unchanged. From the choice of mass parameters and
simple kinematics of the production and decay chain, it is very obvious that
the leading lepton $(\ell_1)$ comes from the two body decay of
heavy $W_R^+ \to \mu^+ + \nu_\mu^c $,
while rest of the leptons originating from the cascade decay chain of
$\nu^c$ and $\tilde\mu^c $ have relatively softer transverse momentum
compared to the $p_T$ of the leading lepton. On the other hand, as the
RH neutrino
mass is increased to a value closer to the $W_R$ mass, the first lepton
becomes softer. However, in this case, the lepton from the decay
$\nu_\mu^c\to e^- \widetilde \mu^c$
merits the highest $p_T$ and will serve as the hardest lepton $(\ell_1)$.

In Fig.~\ref{Xs_A} we show the total signal cross-section $\sigma_s$
(after imposing all the cuts mentioned above) for
the process shown in Eq.~(\ref{csA_approx1}),
as a function of the smuon $\widetilde \mu^c$ mass at the LHC
for 7\,TeV, 10\,TeV and 14\,TeV energies. In each panel,
three curves from top to bottom correspond to $M_{W_{R}} = 1$\,TeV, 1.5\,TeV
and 2\,TeV respectively. We fix the RH neutrino mass
$M_{\nu^c_\mu} = 500 $ GeV and the mixing parameter $g\theta_{eW}=0.2$ for
the present analysis.
Before estimating the possible Standard Model backgrounds to this particular
channel, we would like to discuss the general behaviour of the signal
cross sections.

\begin{itemize}
\item
In all three panels, irrespective of $M_{W_R}$, the $\sigma_s $
first rises with the increases of smuon mass and then becomes almost flat
and finally drops sharply as $M_{\widetilde \mu^c}$ becomes degenerate with
right-handed neutrino mass $M_{\nu^c_\mu}$.

\item
The initial rise of the cross-section with the smuon mass
can be understood from the fact that for lighter smuon mass
$( M_{\widetilde \mu^c} \sim 100-200~{\rm GeV})$, the decay products of smuons
$\widetilde \mu^c \to e^+ \mu^+ j j $ are more collimated and fail
to satisfy our isolation criteria for leptons and jets as
shown in Eq.~(\ref{deltar_cut}). As the smuon mass increases, leptons
and jets which originate from the cascade decay of smuon tend to appear
with larger $\Delta R$, thus satisfying the isolation criteria
as displayed in Eq. (\ref{deltar_cut}).
As a consequence, the $\sigma_s $ for heavier
smuon mass $(M_{\widetilde \mu^c} \lesssim M_{\nu^c_\mu})$ is
significantly larger than for lower smoun mass region.

\item The signal cross secion $\sigma_s$ strongly depends on
$\sqrt{s}$, mass $M_{W_R}$ and off course on $M_{\tilde \mu^c}$.
There is a possibility that the LHC may also run at $\sqrt{s} = 10$\,TeV,
before attaining to its designed $\sqrt{s} = 14$\,TeV. Keeping this
in mind, we decided to provide our observation for $\sqrt{s} = 10$\,TeV
also.  In is very interesting to note that for all the choices of
$M_{W_R}$ and $\sqrt{s}$ the smallest cross-section
always correspond to $M_{\widetilde \mu^c} = 100$\,GeV, while
the largest one correspond to $M_{\widetilde \mu^c} $ which lies
between $400 - 430$\,GeV as shown in Table~\ref{table3}.

\end{itemize}

\begin{table}[thb]
\begin{center}
\begin{tabular}{cccc}
\hline
$M_{W_R}$~&~7\,TeV~&~10\,TeV~&~14\,TeV  \\
\hline
1\,TeV~&~3.2--46.8~&~7.0--100~&~13--178 \\
1.5\,TeV~&~0.3--4.7~&~0.9--14.6~&~1.0--31.7 \\
2\,TeV~&~0.035--0.5~&~0.1--2.3~&~0.4--6.2 \\
\hline
\end{tabular}
\end{center}
\caption[]{The range of minimum and maximum
$\sigma (pp\to e^+ e^- \mu^\pm \mu^\pm jj)$(fb)
at the LHC for $\sqrt{s}=7,\ 10,\ 14\,$TeV and $M_{W_R}=1,\ 1.5,\ 2\,$TeV,
respectively. The corresponding smuon masses are mentioned in the text.
The other parameters are taken as $M_{\nu^c_\mu}=500\,$GeV
and $g\theta_{eW}=0.2$.}\label{table3}
\end{table}

\begin{figure}[t!]
\begin{center}
\includegraphics[width=4.8cm]{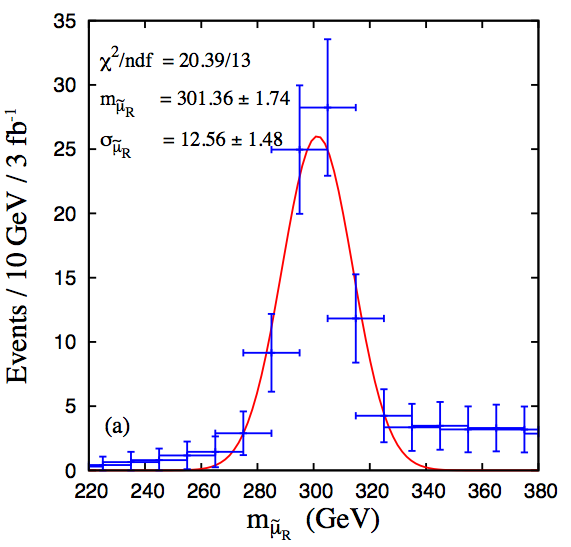}
\includegraphics[width=4.9cm]{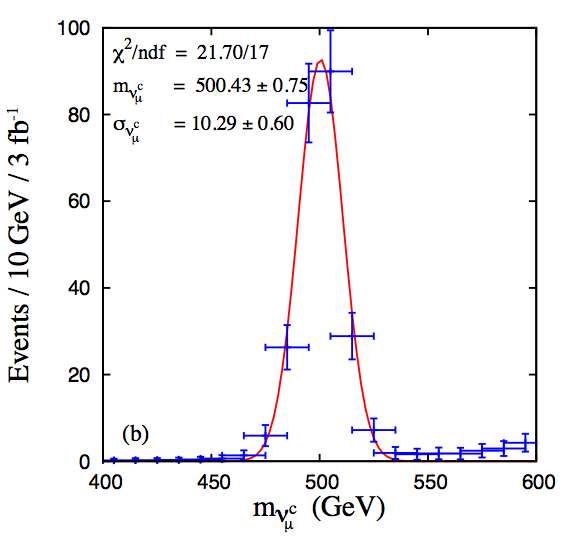}
\includegraphics[width=5cm]{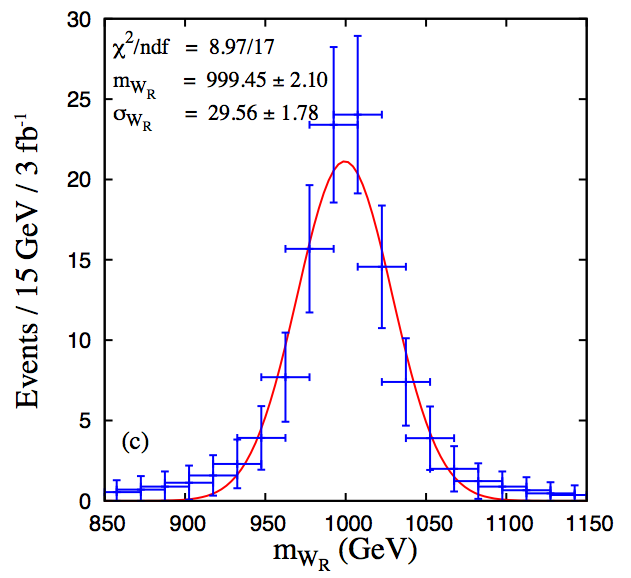}
\caption{Invariant mass distributions
for $M_{\widetilde \mu^c}$, $M_{\nu^c_\mu}$ and $M_{W_R}$
in the $pp\to e^+e^-\mu^\pm\mu^\pm jj$ process
at $\sqrt{s} = 7 $ TeV with $3~{\rm fb}^{-1}$ data
for $M_{\widetilde \mu^c} = 300 $ GeV, $M_{\mu^c_R} = 500 $ GeV and
$M_{W_R} = 1 $ TeV respectively. The error-bars shown are
statistical only for the indicated luminosity. Results of
Gaussian fitting are also shown. }\label{rec_A}
\end{center}
\end{figure}

\bigskip
\noindent{\bf Mass reconstruction}: \ \ The most important feature of our signal events is the effective
reconstruction of all three heavy particle
masses from the final state charged leptons and jets.
We first select two softest leptons (satisfying our selection criteria)
from the four lepton set and then recombine these two leptons
with the two jets to reconstruct the smuon mass,
$M_{jj\ell_3 \ell_4} \approx M_{\tilde \mu^c}$. After the
obtaining the smuon resonance, we attempt to reconstruct the RH
neutrino mass by combining two jets, two softest leptons with one of the
two hardest leptons $\ell_1$ or $\ell_2$.
In this case, we face the complication due to combinatorics with
two choices of pairing for $\ell_{1,2}$ with $M_{jj \ell_3 \ell_4}$.
Finally, $W_R$ can be reconstructed by combining
all four charged leptons and two jets. In Fig.~\ref{rec_A}, we
display the invariant mass distribution for $\widetilde \mu^c, \nu^c_\mu $
and $W_R$ at 7\,TeV LHC. Fitting the mass distribution with a Gaussian, we
get the following values
\begin{eqnarray}
\hspace{-10pt} M_{\widetilde \mu^c}^{\rm fit} = 301.36 \pm 1.74\,{\rm GeV}, \ \
M_{\nu^c_\mu}^{\rm fit} = 500.43 \pm 0.75\,{\rm GeV}, \ \
M_{W_R}^{\rm fit} = 999.45 \pm 2.10\,{\rm GeV}\ ,
\label{mass_fit_A}
\end{eqnarray}
where the input masses considered for this mass reconstruction
procedure are the following
\begin{eqnarray} \label{realmass}
M_{\widetilde \mu^c}^{\rm true} = 300\,{\rm GeV}, \ \
M_{\nu^c_\mu}^{\rm true} =500\,{\rm GeV}, \ \
M_{W_R}^{\rm true} = 1000\,{\rm GeV}\ .
\label{mass_A}
\end{eqnarray}

\begin{table}[thb]
\begin{center}
\begin{tabular}{ccc}
\hline
SM background~&~$\sigma_{0}$ (pb)~&~$\sigma_{\ell^\pm \ell^\pm}$ (fb) \\
\hline
$pp\to b\bar b b \bar b$~&~387.5~&~0.16 \\
$pp\to t\bar t$~&~448~&~0.09 \\
$pp\to Z^0 b \bar b$~&~0.051~&~$3\times10^{-5}$ \\
$pp\to W^\pm W^\pm W^\mp Z^0$~&~$6.7\times10^{-4}$~&~$2\times10^{-5}$ \\
\hline
$\sigma^{\rm total}_{\rm B} $ & & 0.25  \\
\hline
\end{tabular}
\end{center}
\caption[]{The list of leading-order SM backgrounds that could mimic
our signal. $\sigma_{0}$ and $\sigma_{\ell^\pm \ell^\pm}$ are
defined in the text. These numbers correspond to $\sqrt{s}=14\,$TeV.}\label{table2.5}
\end{table}

\noindent{\bf SM backgrounds}: \ \ In principle, there is no intrinsic standard
model background to the $\Delta L = 2 $ processes. However, there some
standard model processes which could mimic our signal
if the missing transverse momentum of neutrinos are balanced.
One of the dominant background is $pp\to b \bar b b \bar b$, followed by
semileptonic decay of all the b-quarks. We generate this background using
with the following basic
cuts $p_T(b)>25\,$ GeV, $|\eta(b)| < 2.5$ and $\Delta R_{bb}>0.4$.
The leading order cross-section is 388 pb at $\sqrt{s} = 14 $ TeV.
After hadronization, one of the $B^0$ or $\overline B^0$ has to oscillate
before decay, in order to get a pair of same-sign dileptons.
The probablity of having
$b\bar b\to e^\pm \mu^\pm, \mu^\pm\mu^\pm$ is about
$P^{b \bar b}_{\ell^\pm \ell^\pm} \approx 2\times 10^{-5}$, as
estimated in~\cite{del Aguila:2007em}. After taking into account the
semileptonic branching ratio $\sim 10\%$ for the other two $b$-quarks we
find this background cross-section $\sim 10^{-1}$~(fb). The other aparently
looking very severe standard model background is $ pp \to t {\bar t} $.
At leading order, the top pair production cross-section
$(\sigma_{t \bar t})$ is 448 pb at the LHC with $\sqrt{s} = 14 $ TeV.
After taking into account the leptonic
branching fraction of two $W$ bosons (from $ t \to b W^+ $) and
$P^{b \bar b}_{\ell^\pm \ell^\pm}$, the rate goes down to $\sim 10^{-1}$ (fb).
Here, we would like to mention that if we take
into account the higher order QCD effects, $\sigma_{t \bar t}$ becomes
$\approx 900$ (pb), which means our final background cross-section from
$t\bar t $ process  may increase atmost by a factor of two.
The other sub-leading standard model
processes which my fake our signal processes are $ pp \to Z b {\bar b}$,
$ pp \to W^\pm W^\pm W^\mp Z^0$ and $ pp \to W^\pm W^\pm W^\mp h$.
In the case of $ pp \to Z b {\bar b}$, process, $Z \to \ell^+ \ell^-,
\ell = e, \mu $ and same sign leptons will come from $b {\bar b}$ pair by
oscillation of one of the $B^0$ meson before decay. As a result of this,
the $\sigma( pp \to Z b {\bar b})$ will be suppressed  by ${\rm Br}(Z \to
\ell^+ \ell^-), \ell = e, \mu$ and $P^{b \bar b}_{\ell^\pm \ell^\pm}$.
The rate for same sign leptons from remaining two processes are
negligibly small. In Table~\ref{table2.5}, we summarize the standard model
background cross-sections, where, $\sigma_{0}$ and
$\sigma_{\ell^\pm  \ell^\pm }$
correspond to the leading order cross-sections before and
after folding with different suppression factors arising from leptonic
branching ratios of $W^\pm, Z $ bosons, semi-leptonic branching ratio of
$b(\bar b)$ quark and finally $P^{b \bar b}_{\ell^\pm \ell^\pm}$
respectively. From this very simple minded exercise, we conclude that
our signal is almost SM background free.

\bigskip
\subsection{$pp\to \mu^\pm e^\pm b\bar b jj$}
The smuon heavier than top quark and with large mixing L-R
mixing $\mathcal{O}(1)$  can lead to this final state.
Here we call the smuon NLSP as $\widetilde\mu$, without definite chirality.
As shown in Table~\ref{table2}, the smuon will dominately decay to $t\bar b$ via the $\lambda'$ coupling in this case.
For this signal topology, we select events with {\it
two same sign different flavoured} (SSDF) charged leptons and four jets.
The cross section for this channel
in the narrow width approximation can be expressed as
\bnq
&&\sigma (pp\to \mu^+ e^+ b\bar b jj) \approx
\sigma ( pp \to W^{+}_R \to \mu^+ \nu^c_{\mu}) \cdot {\rm Br} (\nu^c_{\mu} \to \tilde\mu^- e^+) \cdot
{\rm Br}(\tilde\mu^- \to \bar t b) \cdot {\rm Br}(\bar t \to \bar b jj)\,. \nonumber \\
\enq
The signal also includes the charge conjugated final state $ \sigma (pp\to \mu^- e^- b\bar b jj)$
via intermediate $W^{-}_R$ boson.

Our selection cuts are same as shown in
Eqs.~(\ref{cut:eta})$-$(\ref{ptm_cut}), except for the transverse momentum cut on the jets.
After ordering all four jets according to their $p_T$, we impose following cut on the hardest jet ($j_1$):
\beq
p_T(j_1) > 60~{\rm GeV} \,
\label{ptjhard_cut}
\eeq
and for rest of the jets
\beq
p_T(j_2, j_3, j_4) > 25~{\rm GeV} \ .
\label{ptjrest_cut}
\eeq

\begin{figure}[t]
\begin{center}
\includegraphics[width=6.55cm]{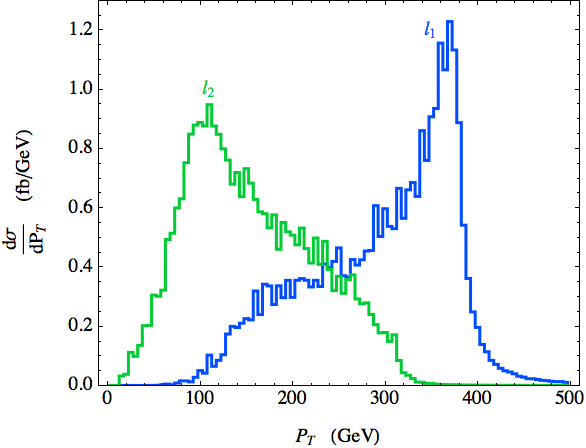} \ \ \
\includegraphics[width=6.25cm]{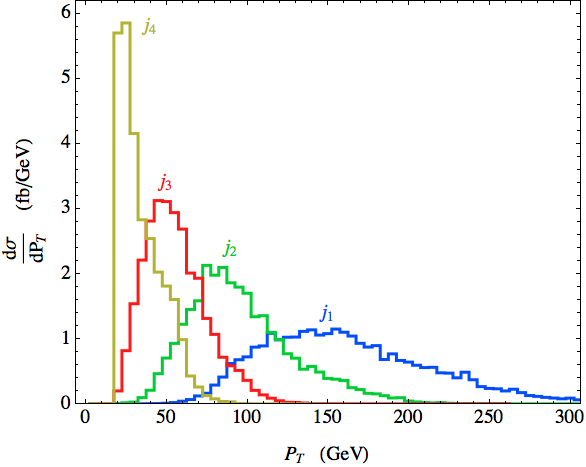}
\caption{$p_T$ distributions of two leptons (left-panel) and all four jets
(right-panel) in the process $pp\to \mu^\pm e^\pm  b {\bar b} jj$ at
$\sqrt{s} = 14$\,TeV.
The leptons and jets are ordered according to their $p_T$ hardness
$(p_T(\ell_1) > p_T(\ell_2)$  and
$(p_T(j_1) > p_T(j_2) > p_T(j_3) > p_T(j_4)$.
The other model parameters are same as in Fig.\ref{distn}.}
\label{ptj_dist_B}
\end{center}
\end{figure}

In Fig.~\ref{ptj_dist_B} we display the
$p_T$ distribution of two leptons and four jets respectively
after ordering them according to their $p_T$. While generating these
distributions, we impose the following cuts on the $p_T$ of
leptons and jets, rest of the cuts remain unchanged,
\begin{eqnarray}
p_T(\ell) > 10\,{\rm GeV}, \ \ \
p^{\rm jets}_T > 15\,{\rm GeV}\,.
\end{eqnarray}
We take the same set of mass parameters as in the previous subsection.
In this case too, the leading lepton comes from the two body decay
$W_R \to \ell_1 + \nu^c_\mu$.
On the other hand, the leading jet $j_1$ mainly comes from the two body decay
of the smuon, while the second hardest jet is produced from the top quark
decay. From the nature of the $p_T$ spectrum of leptons and jets
as shown in Fig.~\ref{ptj_dist_B},
we can justify our choice of $p_T$ cuts (Eqs.~(\ref{ptjhard_cut}) and (\ref{ptjrest_cut})) used in this analysis.

\begin{figure}[t]
\begin{center}
\includegraphics[width=4.9cm]{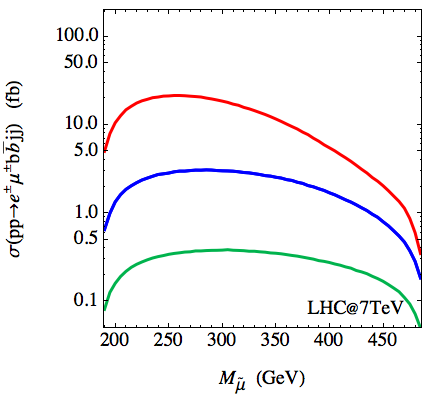}\ \ \
\includegraphics[width=4.9cm]{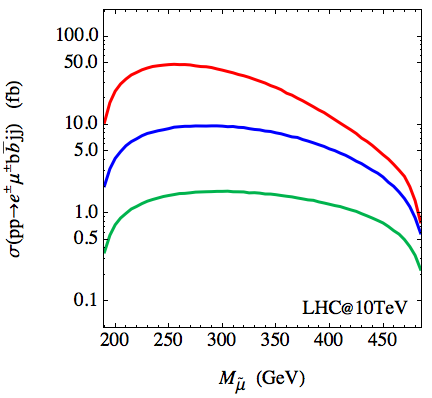}\ \ \
\includegraphics[width=4.9cm]{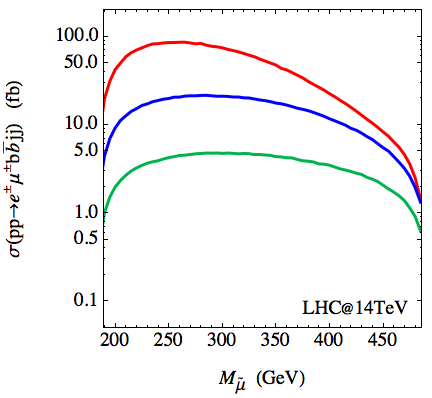}
\caption{Signal cross sections $\sigma (pp\to \mu^-e^-b{\bar b} jj)$
(after all cuts as mentioned in the text) as a function of smuon mass at
the LHC with $\sqrt{s} = 7$\,TeV, 10\,TeV and 14\,TeV. Three curves from
top to bottom in each panel correspond to $M_{W_R} =1$\,TeV,
1.5\,TeV and 2\,TeV respectively. $M_{\nu^c_\mu}$ is kept fixed at 500\,GeV. }
\label{Xs_B}
\end{center}
\end{figure}

In Fig.~\ref{Xs_B} we show the signal cross section (after all cuts
on final state leptons and jets as mentioned above) for this channel
as a function of the smuon mass at the LHC for 7\,TeV, 10\,TeV and 14\,TeV
energies. In each panel, three curves from
top to bottom correspond to $M_{W_R} =1\,$TeV,
1.5\,TeV and 2\,TeV respectively. $M_{\nu^c_\mu} $ is kept fixed at 500\,GeV and $g \theta_{eW}=0.2$,
 the same as in Fig.~\ref{Xs_A}. Comments on the cross sections are in order.

\begin{itemize}
\item
In this case, since we look for $\widetilde \mu^+ \to t {\bar b}$,
we focus on the smuon mass above the top quark threshold, as is displayed
in all three panels of Fig.~\ref{Xs_B}.
As the smuon mass increases, the leptons and jets
originating from the cascade decay of smuon tend to appear with larger
$\Delta R$ between each other, satisfying the isolation criteria shown in
Eq.~(\ref{deltar_cut}).

\item
The signal cross section begins to drop for heavier smuon mass
$(\geq 350~{\rm GeV})$ irrespective of $M_{W_R}$ and
choice of the LHC energy. This is mainly due to
the branching ratio suppression of the $\widetilde \mu^+ \to t {\bar b}$ decay mode,
as can be seen in the right panel of Fig.~{\ref{branching1}}.
Secondly, there is also
the phase space suppression when $M_{\widetilde \mu}$ becomes
close to right-handed neutrino mass $M_{\nu^c_\mu}$.


\item In Table~\ref{table4}, we show the range of signal cross sections for
different values of $M_{\widetilde \mu}$
at the LHC with $\sqrt{s}=7,\ 10,\ 14\,$TeV
and $M_{W_R}=1,\ 1.5,\ 2\,$TeV, respectively.
The other parameters are taken as $M_{\nu^c_\mu}=500\,$GeV
and $g\theta_{eW}=0.2$.
We quote the minimum and maximum values of the signal rate.
For all three choices of $M_{W_R}$ and $\sqrt{s}$ the smallest
cross section always correspond to the value of
$M_{\widetilde \mu} $ which is colse to $M_{\nu^c_\mu}$,
while the largest cross section correspond to $M_{\widetilde \mu}$ lying
between $260 - 300$\,GeV. The signal gets enhanced by more than factor of 2
as the LHC energy increases from 7\,TeV to 10\,TeV,
and by another factor of 2--3 up to 14\,TeV.

\end{itemize}

\noindent{\bf Mass reconstruction}: \ \ We now discuss the mass reconstruction strategy of all three heavy particles
from the final state charged leptons and jets.
From the sample of four jets, the hadronically decaying SM $W$-boson is reconstructed
from pair jets whose invariant mass $(m_{jj})$ is closest to $M_W$. The top
quark is then reconstructed from the reconstructed $W$ and one of the two remaining jets.
We select the one
which gives a invariant mass closest to $M_t$. The smuon mass is reconstructed from
this $M_t$ and with the last jet $M_{\widetilde \mu} \equiv m_{t j}$.
Next, we attempt to reconstruct the
right-handed  neutrino mass by combining
with one of the two leptons $\ell_1$ or $\ell_2$.
In this case, we are facing the combinatorical background with
two choices $m_{t j \ell_1}$, $m_{t j \ell_2}$.
Finally, the $W_R$-boson mass can be reconstructed by combining
all four jets and two charged leptons. We will not explicitly show the reconstruction figure here,
which looks very similar to Fig.~\ref{rec_A}.

\begin{table}[thb]
\begin{center}
\begin{tabular}{cccc}
\hline
$M_{W_R}$~&~7\,TeV~&~10\,TeV~&~14\,TeV  \\
\hline
1\,TeV~&~0.3--21.22~&~0.78--47.9~&~1.4--85 \\
1.5\,TeV~&~0.18--3.04~&~0.58--9.6~&~1.3--21.32 \\
2\,TeV~&~0.05--0.38~&~0.22--1.75~&~0.62--4.73 \\
\hline
\end{tabular}
\end{center}
\caption[]{The $pp\to \mu^\pm e^\pm b \bar bjj$ signal cross sections
(in fb) at the LHC for $\sqrt{s}=7,\ 10,\ 14\,$TeV and $M_{W_R}=1,\ 1.5,\ 2\,$
TeV, respectively. The other parameters are taken as $M_{\nu^c_\mu}=500\,$GeV
and $g\theta_{eW}=0.2$. Here we quote the minimum and maximum values of the
signal rate and the corresponding smuon masses are shown in the text.}
\label{table4}
\end{table}

\begin{table}[thb]
\begin{center}
\begin{tabular}{ccc}
\hline
$\sqrt{s}$~&~$\sigma_{t \bar t W^\pm}$ (fb)~&~$\sigma_{\rm bkg}^{\ell^\pm \ell^\pm}$ (fb) \\
\hline
7\,TeV~&~99~&~1.32 \\
10\,TeV~&~206~&~2.77 \\
14\,TeV~&~377~&~5.05 \\
\hline
\end{tabular}
\end{center}
\caption[]{The dominant SM background $pp\to t\bar t W^\pm$ that could
mimic our signal. $\sigma_{ t \bar t W^\pm }$ correspond to the
production cross-section of $t \bar t W^\pm $ and
$\sigma_{\rm bkg}^{\ell^\pm \ell^\pm} $ represents
cross-section for $  b \bar b j j \mu^\pm e^\pm $ final state before
any cuts. }
\label{table5}
\end{table}

\noindent{\bf SM background}: \ \
In this case, the standard model process which can mimic our signal is
\begin{eqnarray}
p p \to t {\bar t} W^\pm \to b {\bar b} W^+ W^- W^\pm
\to j j b {\bar b} \ell^\pm \ell^{\prime \pm}\,,
\label{smbg.ttw}
\end{eqnarray}
where $\ell, \ell^\prime = e, \mu $.
In our analysis, we do not impose the requirement of $b$ tagging, since
the standard model background also contains $b$-jets, and $b$ tagging would
not improve the signal significance considerably.
The standard model background cross sections from $ pp \to t \bar t W^\pm $
process is shonw in Table.~\ref{table5},
at different LHC energies. We expect this rate would further go down
significantly (by several orders of magnitude)
once we impose our selection criteria on the final state leptons and jets.

We also comment on the other standard model background $pp\to
b\bar b jj$, which has a huge cross section $\sim 10^5\,$pb after
basic cuts.
Taking into account of the oscillation of $b\bar b$ to get same-sign
$e^\pm\mu^\pm$ $P^{b \bar b}_{\ell^\pm \ell^\pm}$ will reduce it
down to the order of $\sim$1\,pb. The cuts on missing energy and the hardest
lepton and jet will further reduce the cross section. Moreover,
in this case, highly energetic $b$-jet will produce charged leptons
which will be very close to the associated $c$-jet, as a result of this,
lepton-jet isolation criteria will play a decisive role in reducing this
background further. Therefore, we conclude that this background will be
also under control. The remaining backgrounds $pp\to W^\pm W^\pm W^\mp Z^0$,
$pp\to W^\pm W^\pm W^\mp h$ and $pp\to jjjj W^\pm W^\pm$ are much smaller~\cite{Perez:2008ha}.

\bigskip

\section{Some Generic Low-energy Constraints}\label{doublebeta}
\subsection{ Neutrinoless double beta decay}
In this model, there are several new contributions to neutrinoless
double beta decay in addition to the usual light neutrino
contribution. The contribution from the RH neutrino exchange as in the non-SUSYLR
models was already discussed~\cite{RNM1}.


In our model, there are two new contributions arising from the $e^c-\tilde{W}_R$ mixing.
The first one is given in left panel of Fig.~\ref{newdb} below. Its contribution to the
effective neutrino mass is given by
\begin{equation}
 m_\nu^{0\nu\beta\beta} \approx \theta_{eW}^2  \left(\frac{M_{W_L}}{M_{W_R}}\right)^4\frac{p_F^2}{m_{\widetilde W_R}} \ ,
\end{equation}
where $p_F\approx50-100\,$MeV is the typical momentum transfer in this process.
For $\theta_{eW}\sim\mathcal{O}(1)$, it is of same
order as the RH neutrino contributions to this process in
non-supersymmetric case.

The second contribution is given in the right panel of Fig.~\ref{newdb} and the corresponding
effective neutrino mass is
\begin{equation}
 m_\nu^{0\nu\beta\beta} \approx \theta_{eW}^2 \left(\frac{\alpha_s}{\alpha}\rule{0cm}{0.5cm}\right) \left(\frac{M_{W_L}}{m_{\widetilde d^c}}\right)^4\frac{p_F^2}{m_{\widetilde g}} \ .
\end{equation}
Note that for this to be consistent with the current limits on the
neutrinoless double beta decay amplitude, we must have
$M_{\tilde{d^c}}, M_{\tilde{G}}\geq 1$ TeV for $g\theta_{eW}\sim
0.2$~\cite{alla}. This however does not constrain the slepton masses which
could still be in the 100 GeV range. Unlike the conventional light
neutrino mass and explicit R-parity violating contributions, these
new contributions lead to RH polarization for the
electron produced in the decay.

\begin{figure}[hbt]
\begin{center}
\includegraphics[width=4.5cm]{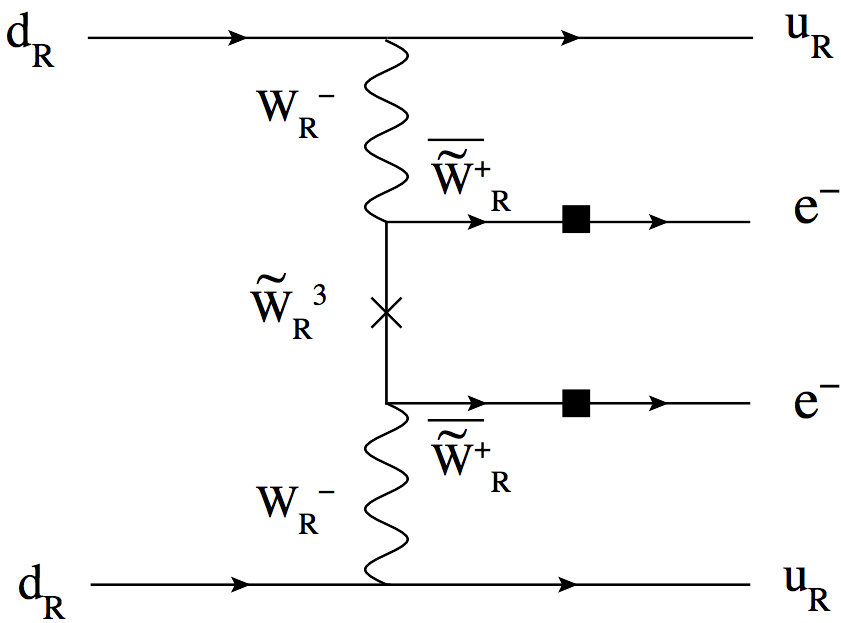} \hspace{1cm}
\includegraphics[width=5cm]{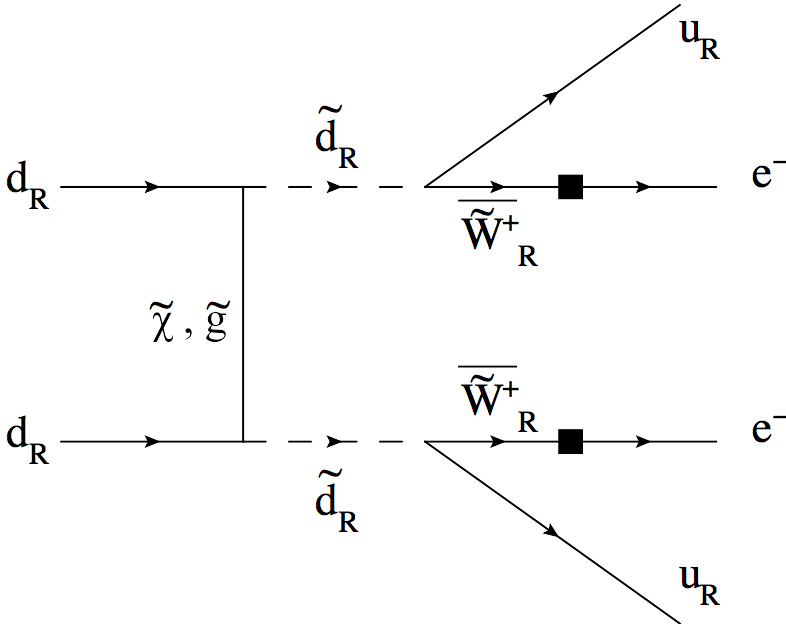}
\caption{New contribution to neutrinoless double beta decay due to
$\widetilde{W}^+_R-e^c$ mixing.}\label{newdb}
\end{center}
\end{figure}

\subsection{ $\pi^0\to e^+e^-$ decay}

This new R-parity violating interaction also has interesting
consequences for rare leptonic decays neutral pion and Kaon
decays. We see from Eq.~(\ref{exotic2}) that via
t-channel $\widetilde{u^c}$ exchange this leads to the process
$\pi^0\to e^+e^-$ with an amplitude given by
$A~\simeq {g^2\theta^2_{eW}}/{M^2_{\tilde{u^c}}}$.
The current PDG bound~\cite{pdg} on this process is ${\rm Br}(\pi^0\to e^+e^-)\leq 6\times 10^{-8}$.
Using the bounds from neutrinoless double beta decay, we predict
that in our model we have ${\rm Br}(\pi^0\to e^+e^-)\leq 10^{-8}$. Note
that if there is mixing in the right-handed charged current of the
same order as the CKM mixings, then we would predict for the $K\to
e^+e^-$ branching ratio at the level about 25 times smaller than
corresponding pion decay. This is about 3 times smaller than the
current PDG quoted bound.
In the LHC search described above, we already restrict ourselves to
this allowed parameter range.

\section{Conclusion}

In summary, we have studied the phenomenology of a class of
minimal SUSYLR models with {\it dynamical R-parity breaking}, i.e., R-parity must
necessarily break in order for parity and gauge symmetry breaking
to occur. This induces a new class of R-parity violating
interactions due to the mixing between $e^c$ and {\footnotesize $\widetilde
W_R^+$}, which are not present in the usual MSSM with explicit or
spontaneous R-parity violation. These interactions lead to a new
contribution to neutrinoless double beta decay which restricts the
squark/gluino masses to be in the TeV range. The model has its
characteristic signature at LHC which consists of final states of
type $e^+e^-\mu^\pm \mu^\pm jj$ or $\mu^\pm e^\pm b\bar b j j$ for
smuon as the NLSP. We estimate the background for this process and
find that for $M_{W_R}$ not far above a TeV, the model should be
testable once LHC reaches its full energy and luminosity.
Incidentally, in this model there is also an upper limit on the
mass of the right-handed $W_R$ boson in the low TeV range for
symmetry breaking to occur. A large part of the mass range could
be accessible even in the early running at the LHC.

\section*{Acknowledgments}
We would like to thank K.S. Babu, B. Bajc, S. Biswas, I.
Gogoladze, T. Han, X. Ji, G. Senjanovi\'c, S. Spinner and J. Zupan
for fruitful discussions. The work of S.L.C. is partially
supported by the US DOE grant DE-FG02-93ER-40762. D.K.G.
acknowledges the hospitality provided by the ICTP High Energy
Group, Trieste, Italy and the Regional Centre for
Accelerator-based Particle Physics (RECAPP), Harish Chandra
Research Institute, Allahabad, India where part of this work was
done. D.K.G. also acknowledges partial support from the Department
of Science and Technology, India under the grant
SR/S2/HEP-12/2006. The work of R.N.M. is supported by the NSF
grant PHY-0968854. The work of Y.Z. is partially supported by the
EU FP6 Marie Curie Research and Training Network “UniverseNet
(MRTN-CT-2006-035863).

\appendix

\section{Fully parity symmetric version}
In this appendix, we consider the full parity symmetric version of
the model. We now keep the $\Delta$ and $\bar\Delta$ multiplets in
our model of Table~\ref{content}. The Yukawa superpotential is given for this
case by the same expression as Eq. (3.1) with two additional
terms: $L^T \tau_2 \Delta L $ and $\mu_\Delta {\rm Tr} \Delta
\bar\Delta$.

 The full potential that is parity symmetric is given below:
\begin{eqnarray}
V_{\rm soft} &=& m_{\widetilde Q}^2 \left( \widetilde Q^\dag
\widetilde Q + \widetilde Q^{c\dag} \widetilde Q^c  \right) +
m_l^2 \left( \widetilde L^\dag \widetilde
L + \widetilde L^{c \dag} \widetilde L^c \right)  \nonumber \\
& + & m_{\Delta}^2 \left[ {~\rm Tr}(\Delta^\dag \Delta)  + {~\rm
Tr}(\Delta^{c\dag} \Delta^c) \right]+ m_{\bar\Delta}^2 \left[
{~\rm Tr}(\bar\Delta^\dag \bar\Delta)
 + {~\rm Tr}(\bar\Delta^{c\dag} \bar\Delta^c) \right] \nonumber \\
& + &\frac{1}{2} \left(  M_{2L} \lambda_L^a \lambda^a_L + M_{2R}
\lambda_R^a \lambda^a_R + M_{1} \lambda_{BL} \lambda_{BL} + M_3
\lambda_g
\lambda_g \right)   \nonumber\\
& + & \widetilde Q^T \tau_2 A^q_i \phi_i \tau_2 \widetilde Q^c +
\widetilde L^T \tau_2 A^\ell_i \phi_i \tau_2  \widetilde L^c + i
A_f \left( \widetilde L_L^T \tau_2 \Delta_L \widetilde L_L +
\widetilde
L_R^{cT} \tau_2 \Delta_R^c \widetilde L_R^c \right)  \nonumber \\
& + &  B_{\Phi\,ab} {\rm Tr}\left( \tau_2 \phi_a^T \tau_2 \phi_b
\right) + B_\Delta  {\rm Tr}\left( \Delta \bar\Delta + \Delta^c
\bar\Delta^c \right)
  + {\rm h.c.} \ .
\end{eqnarray}
The D-term potential as well as the scalar potential can be found
in Refs.~\cite{kuchi, Ji:2008cq}.
The arguments for the existence of the dynamical R-parity breaking
is same as in the parity asymmetric version discussed in sec. 2.
So we do not repeat this discussion here. The only question we
address here is the status of a possible parity symmetric vacuum
\footnote[3]{We thank S. Spinner for raising this point.}
with dynamical R-parity breaking.

First, we would like to understand why the symmetry breaking in the SUSYLR model without
Higgs triplets~\cite{FileviezPerez:2008sx} is not compatible with
the parity symmetry. The point is the LH and RH sneutrinos have opposite $B-L$ charges,
so the D-term potential contributes a negative cross term
\begin{equation}
V_D \sim - \frac{1}{4} g_{BL}^2 \langle\widetilde \nu\rangle^2 \langle\widetilde \nu^{c}\rangle^2  \ ,
\end{equation}
which tends to minimize the potential in the parity conserving $\langle\widetilde \nu\rangle = \langle \widetilde \nu^{c} \rangle$.
This is why the authors of Ref.~\cite{FileviezPerez:2008sx} have to start with parity asymmetric soft mass squared for sneutrinos.

In contrast, the corresponding term in model with Higgs triplets becomes
\begin{equation}
V_D \sim - \frac{1}{4} g_{BL}^2 \left( \langle\widetilde \nu^c\rangle^2 -2 v_R^2 + 2 \bar v_R^2 \rule{0cm}{0.5cm}\right)
\left( \langle\widetilde \nu\rangle^2 -2 v_L^2 + 2 \bar v_L^2 \rule{0cm}{0.5cm}\right) \ ,
\end{equation}
where $\langle \Delta^0 \rangle = v_L$ and $\langle \bar \Delta^0
\rangle = \bar v_L$. According the D-flat condition found out in
Fig.~\ref{correlation}, each bracket is very close to vanishing.
Therefore, such D-term potential does not play significant role in
forcing the vacuum to preserve parity and it is still possible to
start with a symmetric potential. It has been was shown in
\cite{kuchi} that if leptonic Yukawa couplings $Y_\ell$ satisfy
the bound
\begin{eqnarray}
Y^2_\ell~\geq \frac{2f^2(M^2_\Delta-B_\Delta)}{M^2_\Delta} \ ,
\end{eqnarray}
the parity violating minimum is indeed lower than the parity conserving.
By choosing $M_\Delta$ and $B_\Delta$ appropriately, we can
satisfy this bound so that the parity violating and R-parity
violating minimum is the global minimum.

\section{Explicit form of charged fermion mass matrix}

In this appendix, we present the explicit form of charged fermion mass matrix in the SUSYLR model.
The spontaneous R-parity violation induces a mixing between the new chargino {\footnotesize $\widetilde W_R$},
higgsino {\footnotesize $\widetilde{\bar{\Delta}^c}^+$} and the usual electron field.

To see this explicitly, first note that parity violation at the TeV scale
requires spontaneous R-parity breaking at a similar scale, i.e.,
$\langle\widetilde\nu^c_e\rangle \simeq v_R \simeq \bar v_R$.
We can write down the charged fermion mass $\frac{1}{2}
\Psi^T M_{\widetilde C} \Psi + {\rm h.c.}$, in the basis of $\Psi
= [(\widetilde W_R^+, \widetilde{\bar{\Delta}^c}^+, e^{c+}),
(\widetilde W_R^-, \widetilde\Delta^{c-}, e^-)]^T$,
\begin{eqnarray}
M_{\widetilde C} = \left[\begin{array}{cc}
0 & M \\
M^T & 0
\end{array}\right], \ \ \
M=
\left[\begin{array}{ccc}
M_{1/2} & -\sqrt{2} g_R v_R &  0 \\
\sqrt{2} g_R \bar v_R & -\mu_\Delta & 0 \\
g_R \langle\widetilde\nu_e^c\rangle & f \widetilde \langle\nu_e^c\rangle & m_e
\end{array}\right].\
\end{eqnarray}
Following the similar arguments below Eq.~(\ref{charge}), one finds the physical electron field
mass term can be written as
\begin{eqnarray}
\mathcal{L}_m &=& - e m_{e} (\theta_{ee} e^c + \theta_{eW} \widetilde W_R^+ +
\theta_{e\Delta} \widetilde{\bar{\Delta}^c}^+) + {\rm h.c.} \nonumber \\
&\equiv& - e m_{e}\hat e^c  + {\rm h.c.} \ ,
\end{eqnarray}
where $\theta_{ee}, \theta_{eW}, \theta_{e\Delta}$ are
order 1 mixing parameters.

In this model, the role played by $\Delta^c$, $\bar \Delta^c$ Higgses is to give
mass to the RH neutrinos. Meanwhile, their superpartners enter in the above mixing matrix,
but it does not change the generic prediction of large $e^c-${\footnotesize $\widetilde W_R^+$} mixing.

\end{document}